\documentclass[journal]{IEEEtran}
\usepackage{lipsum}
\usepackage[utf8]{inputenc}
\usepackage{amsmath,amsfonts,amssymb,bm}
\usepackage{graphicx}
\usepackage{subcaption}
\usepackage{xcolor}
\usepackage{mathtools}
\usepackage{comment}
\usepackage{textcomp}

\usepackage{multirow}
\usepackage{epstopdf}
\usepackage{flushend}
\usepackage{algorithm}
\usepackage{algpseudocode}
\usepackage{cite}
\usepackage{arydshln}

\newcommand{\bi}{\begin{itemize}}
\newcommand{\ei}{\end{itemize}}

\newcommand{\be}{\begin{enumerate}}
\newcommand{\ee}{\end{enumerate}}
\newcommand{\bd}{\begin{description}}
\newcommand{\ed}{\end{description}}
\newcommand{\bc}{\begin{center}}
\newcommand{\ec}{\end{center}}
\newcommand{\bt}{\begin{tabbing}}
\newcommand{\et}{\end{tabbing}}
\newcommand{\bfig}{\begin{figure}}
\newcommand{\efig}{\end{figure}}
\newcommand{\beq}{\begin{equation}}
\newcommand{\beqarr}{\begin{eqnarray}}
\newcommand{\beqarrn}{\begin{eqnarray*}}
\newcommand{\eeq}{\end{equation}}
\newcommand{\eeqarr}{\end{eqnarray}}
\newcommand{\eeqarrn}{\end{eqnarray*}}
\newcommand{\bflr}{\begin{flushright}\vspace{-0.2in}}
\newcommand{\eflr}{\end{flushright}}
\newcommand{\bsub}{\begin{subequations}}
\newcommand{\esub}{\end{subequations}}
\newcommand{\barr}{\begin{array}}
\newcommand{\earr}{\end{array}}
\newcommand{\nn}{\nonumber}

\def\undb#1{\mbox{\bf{#1}}}

\def\dn{\stackrel{\scriptscriptstyle \triangle}{=}}
\def\BibTeX{{\rm B\kern-.05em{\sc i\kern-.025em b}\kern-.08em
		T\kern-.1667em\lower.7ex\hbox{E}\kern-.125emX}}

\begin{document}
\title{\huge{Secret Key Rate Analysis of RIS-Assisted THz MIMO CV-QKD Systems under Access-Constrained Eavesdropping}}
%\title{RIS-Assisted MIMO CV-QKD: Secret Key Rate Analysis and Phase Optimization Under Constrained Eavesdropping}
\author{Sushil Kumar, Soumya~P.~Dash,~\IEEEmembership{Senior Member,~IEEE}, and George C. Alexandropoulos, \IEEEmembership{Senior Member, IEEE}
% \thanks{This article is an extended version of our earlier work to be presented at IEEE WCNC 2026 \cite{sushilwcnc2026}. Compared to the conference version, this paper provides a significantly enhanced system model incorporating RIS-assisted THz MIMO CV-QKD with detailed quantum channel modeling, rigorous commutation-preserving analysis, extended security evaluation under multiple eavesdropping scenarios, and comprehensive numerical results including optimized RIS phase design.}
\thanks{S. Kumar and S. P. Dash are with the School of Electrical and Computer Sciences, Indian Institute of Technology Bhubaneswar, Argul, Khordha, 752050 India, (e-mails: \{a24ec09010, spdash\}@iitbbs.ac.in).}
\thanks{G. C. Alexandropoulos is with the Department of Informatics and Telecommunications, National and Kapodistrian University of Athens, Panepistimiopolis Ilissia, 16122 Athens, Greece and also with the Department of Electrical and Computer Engineering, University of Illinois Chicago, IL 60601, USA (e-mail: alexandg@di.uoa.gr).}
}
\maketitle

\begin{abstract}
%A reconfigurable intelligent surface (RIS)-aided  wireless communication system is considered in this paper, where the transmitter, Alice, encodes secret keys using Gaussian-modulated coherent states, which are transmitted to the legitimate receiver, Bob, through a composite wireless channel comprising a direct link and a RIS-assisted reflected link. 
A reconfigurable intelligent surface (RIS)-assisted multiple-input multiple-output (MIMO) continuous-variable quantum key distribution (CV-QKD) system operating at terahertz (THz) frequencies, in which a transmitter, Alice, encodes secret keys using Gaussian-modulated coherent states and communicates them to a legitimate receiver, Bob, is considered in this paper. The composite wireless channel, comprising the direct Alice-Bob link and the RIS-assisted reflected link, is modeled as a passive linear Gaussian quantum channel, enabling a unitary dilation that preserves the canonical commutation relations. The security of the considered system is investigated under collective Gaussian entangling attacks by introducing a practical access-constrained eavesdropping model, in which an eavesdropper, Eve, is assumed to access only the environmental modes associated with physically accessible propagation segments. A unified analytical framework is developed to derive the achievable secret key rate (SKR) across all single-segment, pairwise-segment, and full-segment access scenarios, assuming homodyne detection and reverse reconciliation at Bob. Furthermore, an optimization framework is developed to determine the optimal RIS phase configuration matrix and transmitter/receiver beam-splitter parameters that maximize the SKR performance. The resulting optimization problem is efficiently solved using particle swarm optimization. Numerical results are presented to demonstrate the system’s performance with respect to various free parameters. It is showcased that the secrecy performance strongly depends on Eve's accessible propagation segments, with the Alice-RIS channel constituting the most security-critical segment. The proposed access-constrained model achieves substantially higher SKRs than the conventional full-channel purification benchmark, while RIS optimization significantly improves the achievable SKR and extends the secure communication range. This establishes RIS-assisted THz MIMO CV-QKD as a promising solution for next-generation secure wireless networks.
\end{abstract}
%%%%%%%%%%%%%%%%%%%%%%%%%%%%%%%%%%%%%%%%%%%%%%%%%%%%%%%%%%%%%%%%%%%%%%%%%%%%%%%%%%%%%%%%%%%%%%%%%%%%%%%%%%%%%%%
%%%%%%%%%%%%%%%%%%%%%%%%%%%%%%%%%%%%%%%%%%%%%%%%%%%%%%%%%%%%%%%%%%%%%%%%%%%%%%%%%%%%%%%%%%%%%%%%%%%%%%%%%%%%%%%
%%%%%%%%%%%%%%%%%%%%%%%%%%%%%%%%%%%%%%%%%%%%%%%%%%%%%%%%%%%%%%%%%%%%%%%%%%%%%%%%%%%%%%%%%%%%%%%%%%%%%%%%%%%%%%%
%%%%%%%%%%%%%%%%%%%%%%%%%%%%%%%%%%%%%%%%%%%%%%%%%%%%%%%%%%%%%%%%%%%%%%%%%%%%%%%%%%%%%%%%%%%%%%%%%%%%%%%%%%%%%%%
\begin{IEEEkeywords}
Continuous variable quantum key distribution, multiple-input multiple-output, quantum communications, reconfigurable intelligent surfaces, secret key rate.
\end{IEEEkeywords}
%%%%%%%%%%%%%%%%%%%%%%%%%%%%%%%%%%%%%%%%%%%%%%%%%%%%%%%%%%%%%%%%%%%%%%%%%%%%%%%%%%%%%%%%%%%%%%%%%%%%%%%%%%%%%%%
%%%%%%%%%%%%%%%%%%%%%%%%%%%%%%%%%%%%%%%%%%%%%%%%%%%%%%%%%%%%%%%%%%%%%%%%%%%%%%%%%%%%%%%%%%%%%%%%%%%%%%%%%%%%%%%
%%%%%%%%%%%%%%%%%%%%%%%%%%%%%%%%%%%%%%%%%%%%%%%%%%%%%%%%%%%%%%%%%%%%%%%%%%%%%%%%%%%%%%%%%%%%%%%%%%%%%%%%%%%%%%%
%%%%%%%%%%%%%%%%%%%%%%%%%%%%%%%%%%%%%%%%%%%%%%%%%%%%%%%%%%%%%%%%%%%%%%%%%%%%%%%%%%%%%%%%%%%%%%%%%%%%%%%%%%%%%%%
\section{Introduction}
\label{sec:introduction}
The advent of novel technologies for beyond fifth-generation (B5G) and sixth-generation (6G) wireless communication systems is driven by the need to achieve high data rates, ultra-low latency, and high energy and spectral efficiencies \cite{GiPoMe:cm20,9669056}.  
This has led to an increased interest in developing various physical layer techniques, namely extremely multiple-input multiple-output (MIMO)~\cite{9585108,9825647}, operation at terahertz (THz) frequencies~\cite{9766110, 9546670}, integrated sensing and communications (ISAC) \cite{10243495, 10872862}, non-coherent communications \cite{10302402, 8437142, ChenHu2022RISNonCoherent,9689998}, and reconfigurable intelligent surfaces (RIS) \cite{10596064, AlexandropoulosRIS}. Among them, RISs have shown immense potential to revolutionize wireless communications, leveraging vast arrays of response-tunable metamaterial elements that are electronically optimized in almost real time, manipulating electromagnetic waves propagation within the physical wireless channel~\cite{10453467,10255749}. An RIS can strategically direct beams toward designated users, thereby enabling reliable communication with higher data rates, improved signal quality, extended coverage, and reduced interference \cite{10385147}. Moreover, implementing RIS technology enables robust communication links for systems dominated by non-line-of-sight (NLoS) signal propagation~\cite{Alexandropoulos2023RISEnabled}. This has led to the usage of RIS to improve the performance of MIMO, ISAC, and index modulation systems, as well as satellite networks~\cite{10054402, 9698029, 9802114, 10400440, 10247149,10388479, 10287142}.

Another crucial aspect and requirement for next-generation communication systems is improving the security and privacy of transmitted data. Traditional higher-layer encryption techniques based on Public-key cryptosystems, such as Rivest-Shamir-Adleman (RSA) are vulnerable to quantum algorithms such as Shor’s algorithm when sufficiently large fault-tolerant quantum computers become available\cite{weedbrook2012gaussian, manzalini2020quantum}. Moreover, the development of quantum computers has made the use of algorithms for physical-layer encryption, such as Diffie-Hellman, inefficient, as they can yield solutions to computationally hard discrete logarithm problems in a short time \cite{diffie1976new}. To overcome these challenges and enhance data privacy, quantum key distribution (QKD), a technique that leverages the principles of quantum superposition and entanglement, has been proposed in the literature as a versatile secure communication technique~\cite {8865103, 9845436, 9552894}.

QKD ensures the secure transmission of secret keys between two authenticated users, namely Alice and Bob, even in the presence of a potential eavesdropper, Eve. The literature on QKD classifies the available techniques into two categories, i.e., discrete-variable QKD (DV-QKD) and continuous-variable QKD (CV-QKD) \cite{8865103, cv_dv_qkd_Scarani_2009}. 
The DV-QKD approach relies on sources and detectors specifically designed for single photons and encrypts confidential key information using the polarization or phase of a single photon. The secure key produced by the DV-QKD approach is guaranteed by the no-cloning theorem of quantum physics, preventing the possibility of making perfect replicas of non-orthogonal quantum states without introducing detectable noise \cite{9102386, 9799745}. In contrast, a secret key is encoded using the quadratures of continuous variable Gaussian quantum states with the CV-QKD technique, and its security is ensured by Heisenberg's uncertainty principle \cite{weedbrook2012gaussian, 8732438}. Owing to higher hardware compatibility with classical communication systems and the ability to offer superior key rates and overall performance in noisy environments, as well as ease of implementation in higher frequency ranges, the CV-QKD approach becomes a natural choice for wireless systems over the DV-QKD one~\cite{lodewyck2007quantum, cvqkd_ppf_2024}.

Most QKD-enhanced wireless systems focus on applications that require point-to-point communications, such as satellite-to-earth, inter-satellite, inter-building, and free-space maritime channels \cite{pirandola2021limits, pan2020secretOptics, pirandola2021satellite, gariano2017engineering, gariano2018trade,sushil_2wayCVQKD_hybridnoise}, which are traditionally achieved via free-space optical (FSO) links~\cite{8920091}. However, it has been studied that THz communication is often preferred over FSO due to its better resilience to weather, NLoS capabilities, easier deployment (with less strict alignment), and ability to handle high data rates over short distances \cite{mmWaveThzMIMO_pirandola_10104156}.
This has led to several studies for terrestrial and inter-satellite wireless systems appearing in the literature considering the usage of the THz band for secure communications using the CV-QKD technique~\cite{ottaviani2020terahertz,inter_satellite_QKD_thz_Wang_2019, satellite2ground_CVQKD_10415457}. However, these studies have reported low secret key rates (SKRs) and short key transmission distances due to several factors that degrade the channel within the considered frequency spectrum. These limitations have been shown to be overcome to some extent by the integration of MIMO with CV-QKD \cite{2_ref_paper,sushil_2wayCVQKD_11129674}. MIMO FSO with DV-QKD has been studied in~\cite{sushil_ICC2026}. On another front, RISs have also been lately integrated with QKD. The authors in \cite{Qkd_with_ris_Sayeed:22} introduced the idea of reduction of reflection loss with the use of RIS, and the authors in~\cite{11152330_sushil_RIS} studied channel estimation and RIS to aid QKD-based quantum communication in a THz MIMO system. However, existing studies on RIS-assisted THz MIMO CV-QKD mainly focus on channel enhancement, channel estimation, and SKR optimization under a fixed eavesdropping model. The security impact of Eve's segment-wise access to the environmental modes associated with the direct Alice-Bob, Alice-RIS, and RIS-Bob propagation links remains largely unexplored. In particular, there is no unified access-constrained framework that compares single-segment, pairwise-segment, and full-segment Eve access in RIS-assisted THz MIMO CV-QKD systems~\cite{11152330_sushil_RIS}.

Motivated by the above research gap, this paper investigates an RIS-assisted THz MIMO CV-QKD system in which Alice and Bob are connected through both a direct path and an RIS-assisted reflected path. Unlike conventional RIS-assisted wireless systems, where the RIS is typically optimized to enhance Bob's received signal strength, the objective in CV-QKD is to maximize the secret key rate (SKR), which depends on both mutual information between Alice and Bob and Eve's quantum side information. Hence, a RIS configuration that improves Bob's channel may also increase Eve's information if the corresponding environmental modes are accessible to her.
To capture this effect, we propose an access-constrained environmental-mode eavesdropping framework. The RIS-assisted THz MIMO channel is decomposed into three propagation segments: the direct Alice-Bob link, the Alice-RIS link, and the RIS-Bob link. Eve is not assumed to voluntarily restrict her attack to a single segment; rather, she is assumed to collect all environmental modes that are physically accessible from her feasible deployment region. The global purification model, where Eve controls the purification of the entire Alice-Bob channel, is retained as a conservative benchmark. However, in practical RIS-assisted THz deployments, spatial separation of propagation corridors, physical protection, blockage, restricted deployment regions, and directional propagation may prevent Eve from simultaneously accessing all environmental modes~\cite{11078147}.
Thus, this work extends our preliminary study in \cite{sushilwcnc2026}, which considered RIS phase optimization under an end-to-end eavesdropping model. The present work develops a segment-resolved access-constrained security analysis. We construct the covariance matrices and Holevo information terms for arbitrary Eve access sets over the direct, Alice-RIS, and RIS-Bob environmental modes. This enables us to compare single-segment, pairwise-segment, and full-segment access cases and to identify the propagation segment that dominates the SKR degradation.

As a result, although RIS-assisted THz MIMO CV-QKD has recently attracted considerable attention, prior work has primarily focused on end-to-end channel optimization or a single eavesdropping model~\cite{11152330_sushil_RIS}. As a result, the propagation segment that is most vulnerable under physically constrained eavesdropping remains unidentified, while a unified covariance-based SKR framework for arbitrary combinations of accessible environmental modes remains unavailable. In addition, the influence of RIS design and placement on the worst-case access-constrained SKR has not been systematically explored.
The major contributions of this paper are summarized as follows:
\begin{itemize}

\item We develop an access-constrained environmental-mode eavesdropping framework for RIS-assisted THz MIMO CV-QKD systems. The proposed framework models Eve's side information based on the propagation segments physically accessible to her.

\item  We decompose the RIS-assisted THz MIMO channel into the direct Alice-Bob, Alice-RIS, and RIS-Bob propagation segments, and derive analytical SKR expressions for all feasible Eve access sets, including single-segment, pairwise-segment, and full-segment access scenarios.

\item We formulate a robust RIS optimization problem to jointly optimize the RIS phase shifts and beam-splitting parameters to maximize the achievable SKR across all feasible Eve access sets, and efficiently solve the resulting non-convex problem using particle swarm optimization (PSO).

\item We identify the security-critical propagation segments of RIS-assisted THz MIMO CV-QKD. In particular, we show that Eve's access to the Alice-RIS segment consistently results in the largest SKR degradation and dominates the secrecy performance whenever it is included in the compromised access set.

\item We present extensive numerical results to quantify the impact of Eve's access set, MIMO dimension, RIS size, transmission distance, modulation variance, detector noise, and reconciliation efficiency on the achievable SKR. The results demonstrate that the proposed PSO-based RIS design consistently outperforms random-phase RIS configurations and non-RIS transmission, while access-constrained eavesdropping yields substantially higher SKRs than the conventional global-purification 
\end{itemize}

The rest of the paper is organized as follows. The system model of the RIS-assisted MIMO CV-QKD system, the transmission of secret keys between Alice and Bob, and the considered eavesdropping models are detailed in Section II. The SKR analysis under both localized and global eavesdropping models, along with the PSO algorithm, is presented in Section III. Finally, Section IV includes our numerical results, followed by concluding remarks in Section V.

\textit{Notation:} Boldface letters such as $\undb{A} \in {\mathbb{C}}^{M \times N} $ and $\undb{a} \in {\mathbb{C}}^{M \times 1} $ represent matrices and vectors, respectively. The $\textbf{A}^\dagger$ symbol stands for the conjugate transpose of $\textbf{A}$, the $\textbf{A}^T$ symbol stands for the transpose, the $\textbf{A}^{+}$ symbol represents Moore–Penrose pseudo-inverse, and $\textbf{A} \otimes \textbf{B}$ indicates the tensor product between $\textbf{A}$ and $\textbf{B}$. The notations $\boldsymbol{1}_{M\times N}$ and $\boldsymbol{0}_{M\times N}$ represent a matrix consisting of all ones and all zeros, respectively, and $\jmath\triangleq\sqrt{-1}$ is the imaginary unit. $\textbf{I}_M$ represents a $M \times M$ identity matrix, and $\text{diag}(\undb{a})$ returns a $M \times M$ diagonal matrix with the elements of $\undb{a}$ on its principal diagonal. $a^*$ represents the conjugate of $a$. Notation $\langle X \cdot Y\rangle$ indicates the quantum correlation between $X$ and $Y$, while $\mathcal{N} (\boldsymbol{\mu},\boldsymbol{\sigma^2})$ represents the real multivariate Gaussian distribution, where $\boldsymbol{\mu}$ is the mean vector and $\boldsymbol{\sigma^2}$ is the covariance matrix. $[\_,\_ ]$ is the canonical bosonic commutation, $|\cdot|$ denotes the magnitude operator, and the operator $\text{eig} (\cdot)$ computes the eigenvalues of a matrix. Furthermore, $\hat{Q}$ denotes the operator (such as annihilation and creation) acting on the signal mode $Q$.

\begin{figure*}[!t]
    \centering
    \includegraphics[width=18cm,height=6.5cm]{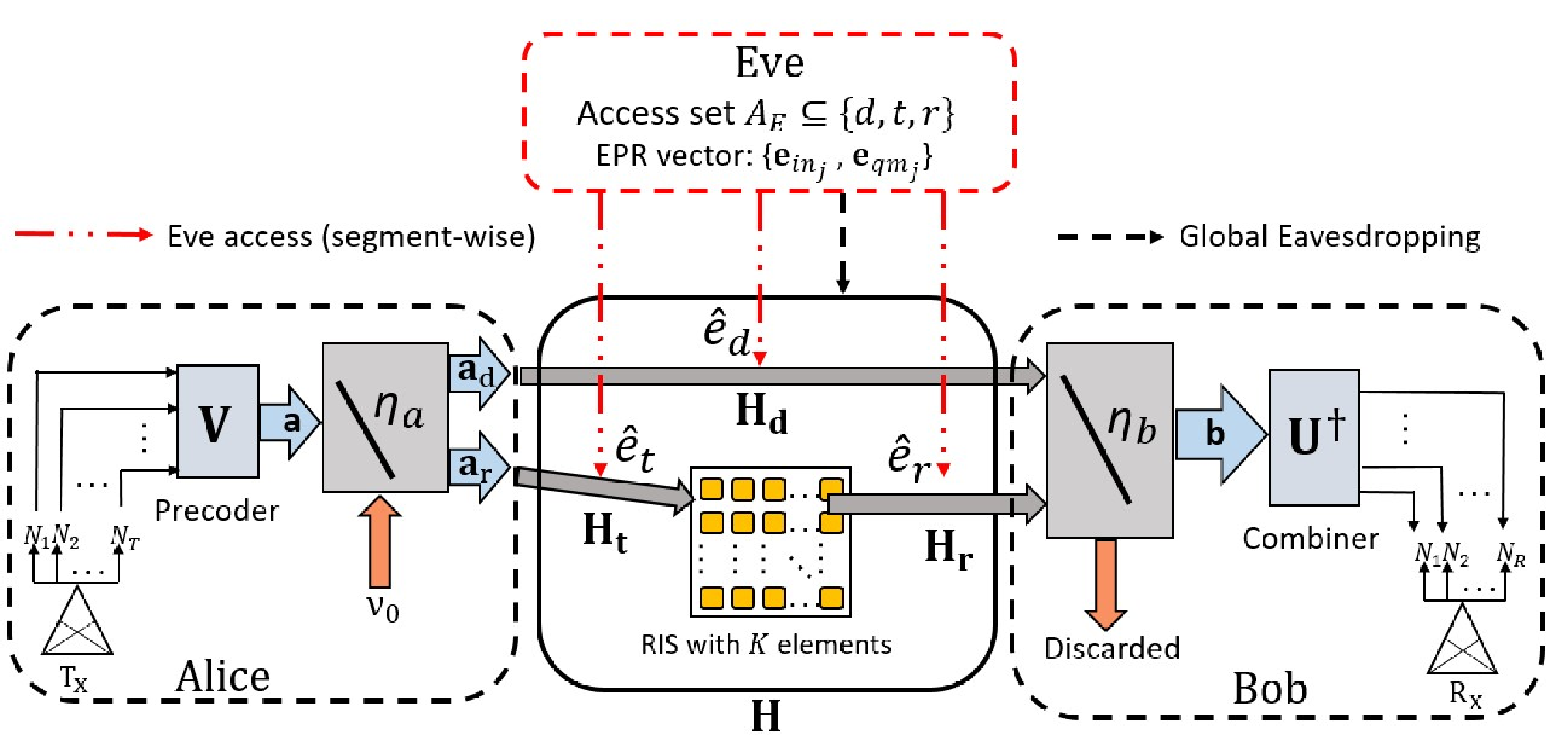} \vspace{-0.6cm}
    \caption{The considered RIS-assisted MIMO CV-QKD system with access-constrained segment-wise eavesdropping.}
    % \caption{The considered RIS-assisted MIMO CV-QKD wireless communication system model with access-constrained segment-wise eavesdropping.}
    \label{f1}
\end{figure*}
%%%%%%%%%%%%%%%%%%%%%%%%%%%%%%%%%%%%%%%%%%%%%%%%%%%%%%%%%%%%%%%%%%%%%%%%%%%%%%%%%%%%%%%%%%%%%%%%%%%%%%%%%%%%%%%
%%%%%%%%%%%%%%%%%%%%%%%%%%%%%%%%%%%%%%%%%%%%%%%%%%%%%%%%%%%%%%%%%%%%%%%%%%%%%%%%%%%%%%%%%%%%%%%%%%%%%%%%%%%%%%%
\section{System and Channel Models}

We consider an RIS-assisted MIMO CV-QKD system as shown in Fig. \ref{f1}, where the transmitter, Alice, and the receiver, Bob, are equipped with $N_T$ and $N_R$ antennas, respectively.
% \begin{figure*}[!t]
%     \centering
%     \includegraphics[width=18cm,height=6.5cm]{W6_SM.eps}
%     \caption{The considered RIS-assisted MIMO CV-QKD wireless communication system model under eavesdropping attacks.}
%     \label{f1}
% \end{figure*}
The RIS consists of $K$ passive reflecting elements configured to adjust the phases of the incident and transmitted electromagnetic signals~\cite{AlexandropoulosRIS}. This communication is also aided by the direct LoS path between the transceiver pair. 

\subsection{Channel Model}
The effective MIMO channel between Alice and Bob, denoted by $\mathbf{H}\in \mathbb{C}^{N_R\times N_T}$, can be mathematically expressed as follows~\cite{10453467}:
\beq
\mathbf{H} \triangleq \mathbf{H}_\text{d} + \mathbf{H}_\text{r} \mathbf{\Phi} \mathbf{H}_\text{t} \, ,
\label{eq1}
\eeq
where $\mathbf{\Phi} \triangleq \text{diag} \left(e^{\jmath \phi_{1}},...,e^{\jmath \phi_{K}}\right)$ and $\phi_k$ is the phase shift introduced by the $k$-th element of the RIS. Furthermore, $\mathbf{H}_\text{d}\in {\mathbb{C}}^{N_R \times N_T}$, $\mathbf{H}_\text{t} \in \mathbb{C}^{K\times N_T}$, and $\mathbf{H}_\text{r}\in \mathbb{C}^{N_R \times K}$, are the direct LoS channel matrix between the legitimate transceiver pair, the channel matrix between Alice and RIS, and the channel matrix between RIS and Bob, respectively, which are expressed for the THz system as follows:
\beqarr
&& \! \! \! \! \! \! \! \! \! \!
\! \! \! \! \! \! \! \! \! \! \mathbf{H}_\text{d} \triangleq
\sum_{\ell_d=1}^{L_d} \sqrt{\delta_{\ell_d}}
e^{\jmath 2 \pi f_c \tau_{\ell_d}} 
\mathbf{h}_{N_R}\left( \theta_{\ell_d}^{R_X} \right)\mathbf{h}^\dagger_{N_T}
\left(\theta_{\ell_d}^{T_X} \right) \, , \nn \\
&& \! \! \! \! \! \! \! \! \! \!
\! \! \! \! \! \! \! \! \! \! \mathbf{H}_\text{t} \triangleq
\sum_{\ell_t=1}^{L_t} \sqrt{\delta_{\ell_t}}
e^{\jmath 2 \pi f_c \tau_{\ell_t}} \mathbf{h}_{\text{RIS}} 
\left(\varphi,\theta_{\ell_t}^{\text{RIS}}\right)
\mathbf{h}^\dagger_{N_T}
\left( \theta_{\ell_t}^{T_X} \right) \, , \nn \\
&& \! \! \! \! \! \! \! \! \! \!
\! \! \! \! \! \! \! \! \! \! \mathbf{H}_\text{r} \triangleq
\sum_{\ell_r=1}^{L_r} \sqrt{\delta_{\ell_r}}
e^{\jmath 2 \pi f_c \tau_{\ell_r}}
\mathbf{h}_{N_R} \left( \theta_{\ell_r}^{R_X} \right) 
\mathbf{h}^\dagger_{\text{RIS}}
\left( \varphi,\theta_{\ell_r}^{\text{RIS}} \right) \, .
\label{eq2}
\eeqarr
In these expressions, $L_d, L_t$, and $L_r$ denote the number of signal propagation paths in the wireless channels $\mathbf{H}_\text{d}$, $\mathbf{H}_\text{t}$, and $\mathbf{H}_\text{r}$, respectively. Further, $f_c$ is the carrier frequency, and $\tau_{\ell_d}$, $\tau_{\ell_t}$, $\tau_{\ell_r}$ and $\delta_{\ell_d}$, $\delta_{\ell_t}$, $\delta_{\ell_r}$ are the propagation delays and path losses corresponding to $\mathbf{H}_\text{d}$, $\mathbf{H}_\text{t}$, and $\mathbf{H}_\text{r}$, respectively. Moreover, $\theta_{\ell_d}^{T_X}$, $\theta_{\ell_t}^{T_X}$, and $\theta_{\ell_r}^{\text{RIS}}$ are the angles of departure (AoD) and $\theta_{\ell_d}^{R_X}$, $\theta_{\ell_t}^{\text{RIS}}$, and $\theta_{\ell_r}^{R_X}$ are the angles of arrival (AoA) of the $\ell_d$-th, $\ell_t$-th, and $\ell_r$-th multipath from the transmitter's and the receiver's uniform linear arrays (ULAs), respectively. The antenna elements in both ULAs are uniformly placed in a single dimension such that the inter-element spacing is maintained at $d_a$. This results in the array response vector in (\ref{eq2}), i.e., $\mathbf{h}_{N_T} \left( \text{or } \mathbf{h}_{N_R} \right)$, are given as follows:
\beq
\mathbf{h}_{N} \left( \theta \right) \triangleq \frac{1}{\sqrt{N}} 
\left[1, e^{\jmath \frac{2\pi}{\lambda_c}
d_a \sin \theta} , \cdots ,
e^{\jmath \frac{2\pi}{\lambda_c}
d_a \left( N-1 \right) \sin \theta} \right]^T ,
\label{eq3}
\eeq
where $N \in \{N_T,N_R\}$ and $\lambda_c \triangleq c/f_c$, with $c$ being the speed of electromagnetic signals (i.e., light). Furthermore, $\mathbf{h}_{\text{RIS}}\in \mathbb C^{K\times 1}$ in (\ref{eq2}) is the response vector of the passive ULA of the RIS \cite{path_los_RIS_Ellingson_2021}, \cite{ris_assisted_mmwave_mimo_9918631} and is given as
\beqarr
&& \! \! \! \! \! \! \! \! \!
\! \! \! \! \! \! \! \! \! \!
\mathbf{h}_{\text{RIS}} 
\left( \varphi, \theta_{\ell}^{\text{RIS}} \right)
= \frac{1}{\sqrt{K}} \left[e^{\jmath\frac{2\pi}{\lambda_c}
\left(\vartheta_X^{\varphi, \theta_{\ell}^{\text{RIS}}} 
+ \vartheta_Y^{\varphi, \theta_{\ell}^{\text{RIS}}} \right)}, \cdots \right. \nn \\
&& \qquad \left. \cdots ,
e^{\jmath\frac{2\pi}{\lambda_c}
\left( \left( K_X-1 \right) \vartheta_X^{\varphi, \theta_{\ell}^{\text{RIS}}}
+ \left( K_Y - 1 \right) \vartheta_Y^{\varphi, \theta_{\ell}^{\text{RIS}}} \right)} \right] ,
\label{eq4}
\eeqarr
where, for $\ell \in \left\{\ell_t , \ell_r \right\}$, it holds:
\beqarr 
\vartheta_X^{\varphi, \theta_{\ell}^{\text{RIS}}}
\! \! \! \! &=& \! \! \! \! d_X \cos\left(\varphi\right)
\sin\left(\theta_{\ell}^{\text{RIS}} \right) ,\nn\\
\vartheta_Y^{\varphi, \theta_{\ell}^{\text{RIS}}}
\! \! \! \! &=& \! \! \! \! d_Y \sin\left(\varphi\right)
\sin \left(\theta_\ell^{\text{RIS}} \right) \,.
\label{eq5}
\eeqarr
Here, the elements of the RIS are considered to be arranged along a 2-dimensional structure \cite{Alexandropoulos2023RISEnabled, ris_channel} with $K_X$ and $K_Y$ reflecting elements along the horizontal and the vertical axes, respectively, implying that $K_X K_Y=K$, with the separation between the elements in the corresponding axes being denoted by $d_X$ and $d_Y$. Furthermore, the path losses in (\ref{eq2}) are expressed as follows:
\beq
\delta_{j,\ell} \! = \!
\begin{cases} \! \! 
    \left( \frac{\lambda_c}{4\pi d_{j,\ell}} \right)^2 
    \! \! G_{T_X} G_{R_X} 10^{-0.1 \rho d_{j,\ell}} \, , j \in \left\{ d, t, r \right\} , \ell=1 \\
    \! \varsigma \xi_\ell \left( \frac{\lambda_c}
    {4\pi d_{j,\ell}} \right)^2 \! \! G_{T_X} G_{R_X}
    10^{-0.1 \rho d_{j,\ell}}
    \! , \ell \in \left\{\ell_d, \ell_t, \ell_r \neq 1 \right\} \! ,
\end{cases}
\label{eq6}
\eeq
where $d_{j,\ell}$ is the smallest path length,  $\xi_\ell$ is the Fresnel reflection coefficient of the $\ell$-th multipath component, $\varsigma$ is the Rayleigh roughness factor, and $\rho$ (in dB/km) is the atmospheric absorption loss. Moreover, Alice's and Bob's ULAs gains are $G_{T_X} = N_TG_a$ and $G_{R_X} = N_RG_a$, respectively, where $G_a$ is the gain of each component of the transmitter and receiver antennas, and for the Alice-RIS and RIS-Bob links, the array gain associated with the RIS side is modeled through the number of reflecting elements $K$, while the Alice and Bob array gains are given by $N_TG_a$ and $N_RG_a$, respectively.
%%%%%%%%%%%%%%%%%%%%%%%%%%%%%%%%%%%%%%%%%%%%%%%%%%%%%%%%%%%%%%%%%%%%%%%%%%%%%%%%%%%%%%%%%%%%%%%%%%%%%%%%%%%%%%%%%%%%%%%%%%%%%%%%%%%%%%%%%%%%%%%%%%%%%%%%%%%%%%%%%%%%%%%%%%%%%%%%%%%%%%%%%%%%%%%%%%%%%%%%%%%%%%%%%%%%%%%%%%%%%%%%%%%%%%%%%%%%%%%%%%%%%%%%%%%%%%%%%%%%%%%%%%%%%%%%%%%%%%%%%%%%%%%%%%%%%%%%%%%
\subsection{Secret Key Generation, Transmission, and Reception}
In the considered MIMO wireless communication system, Alice aims to establish a secure key with Bob. To this end, Alice employs a Gaussian-modulated CV-QKD scheme, generating two independent zero-mean Gaussian random vectors
$\undb{X}_{\text{Alice}}, \undb{P}_{\text{Alice}} \sim {\mathcal{N}} \left( \mathbf{0}_{N_T}, V_s \undb{I}_{N_T} \right)$ corresponding to the position and momentum quadratures, respectively \cite{2_ref_paper, QC_book}. Using these quadratures, Alice generates a set of coherent states, ${a}_n = X_{\text{Alice},n} + \jmath P_{\text{Alice},n}$, for $n = 1, \ldots, N_T$, which are transmitted to Bob through the wireless propagation medium. After propagation through the wireless medium, the received quantum signal at Bob is modeled as the coherent combination of the fields arriving through the direct and RIS-assisted paths. This operation is represented by a beam splitter (BS$_b$) with transmissivity $\eta_b$, yielding the received quantum signal expression:
\beq
\hat{\mathbf{b}} \triangleq \sqrt{\eta_b}\,\hat{\mathbf{b}}_\text{d} + \sqrt{1-\eta_b}\,\hat{\mathbf{b}}_\text{r} ,
\label{eq7}
\eeq
where $\hat{\mathbf{b}}_\text{d}$ and $\hat{\mathbf{b}}_\text{r}$ represent the signals arriving from the direct and RIS-assisted channels, respectively. The signal to the direct path and the RIS-assisted path are expressed as follows:
\bsub
\beq
\hat{\mathbf{b}}_\text{d}\triangleq
\mathbf{H}_\text{d}  \hat{\mathbf{a}}_\text{d}
+ \mathbf{N}_\text{d} \hat{\mathbf{e}}_\text{d} \, ,
\label{eq8a}
\eeq
\beq
\hat{\mathbf{b}}_\text{r}\triangleq\left(\mathbf{H}_\text{r} \mathbf{\Phi}\mathbf{H}_\text{t}\right)\hat{\mathbf{a}}_\text{r}
+\mathbf{H}_\text{r}\mathbf{\Phi}\mathbf{N}_\text{t}\hat{\mathbf{e}}_\text{t}
+\mathbf{N}_\text{r} \hat{\mathbf{e}}_\text{r} \, ,
\label{eq8b}
\eeq
\esub
respectively, where $\hat{\mathbf{e}}_\text{d}$, $\hat{\mathbf{e}}_\text{t}$, and $\hat{\mathbf{e}}_\text{r}$ denote mutually independent bosonic environmental input modes associated with the losses along the direct Alice-Bob, Alice-RIS, and RIS-Bob propagation segments, respectively. They satisfy $\left[\hat{\mathbf{e}}_j,\hat{\mathbf{e}}_j^\dagger\right]=\mathbf{I_{N_j}}, \quad \left[\hat{\mathbf{e}}_i,\hat{\mathbf{e}}_j^\dagger\right]=\mathbf{0},\; i\neq j$, where $i,j\in \left\{d,t,r\right\}$ and $N_j\in \left\{\min(N_T,N_R),\min(N_T,K),\min(K,N_R)\right\}$. To generate independent propagation modes for the direct and RIS-enabled channel paths, Alice first splits the transmitted signal using a beam splitter (BS$_{a}$) with transmissivity $\eta_a$ as follows:
\bsub
\beq
\hat{\mathbf{a}}_\text{d}\triangleq\sqrt{\eta_a}\, \hat{\mathbf{a}}
+ \sqrt{1-\eta_a}\,\hat{\mathbf{v}}_0 \, ,
\label{eq9a}
\eeq
\beq
\hat{\mathbf{a}}_\text{r}\triangleq-\sqrt{1-\eta_a}\, \hat{\mathbf{a}}
+ \sqrt{\eta_a}\,\hat{\mathbf{v}}_0 \, ,
\label{eq9b}
\eeq
\esub
where $\hat{\mathbf{v}}_0$ is a vacuum mode vector and $\hat{\mathbf{a}}$ is the Alice transmitted mode vector. These operators satisfy the canonical bosonic commutation relations:
\beq\left[\hat{\mathbf{a}}, \hat{\mathbf{a}}^\dagger\right]=\mathbf{I}_{N_T}, \left[\hat{\mathbf{v}}_0,\hat{\mathbf{v}}_0^\dagger\right]=\mathbf{I}_{N_T}, \left[\hat{\mathbf{a}},\hat{\mathbf{v}}_0^\dagger \right]=\mathbf{0}_{{N_T}\times {N_T}}.
\label{eq10}
\eeq
Furthermore, each propagation segment is modeled as a physically realizable passive Gaussian channel arising from a unitary interaction between the signal and the corresponding environmental mode, which is given as $\mathbf{H}_{j}\mathbf{H}_{j}^\dagger + \mathbf{N}_{j}\mathbf{N}_{j}^\dagger=\mathbf{I}_{N_j}\, ,\forall\, j\in\{d,t,r\}$. Here, $\mathbf{N}_{j}$ denotes the environmental coupling matrices which characterize the interaction between the signal and the corresponding environmental mode $\hat{\mathbf{e}}_j$, $\forall j\in \{d,t,r\}$. Since the signal and environmental modes originate from a single global unitary transformation, the received operator $\hat{\mathbf{b}}$ at Bob satisfies the canonical bosonic commutation relations:
\beq
\left[\hat{\mathbf{b}}, \hat{\mathbf{b}}^\dagger\right] = \mathbf{I}_{N_R}.
\label{eq11}
\eeq
The segment-wise environmental modes in \eqref{eq8a} and \eqref{eq8b} serve as the basis for the access-constrained eavesdropping model introduced in the next subsection.
%%%%%%%%%%%%%%%%%%%%%%%%%%%%%%%%%%%%%%%%%%%%%%%%%%%%%%%%%%%%%%%%%%%%%%%%%%%%%%%%%%%%%%%%%%%%%%%%%%%%%%%%%%%%%%%%%%%%%%%%%%%%%%%%%%%%%%%%%%%%%%%%%%%%%%%%%%%%%%%%%%%%%%%%%%%%%%%%%%%%%%%%%%%%%%%%%%%%%%%%%%%%%%%%%%%%%%%%%%%%%%%%%%%%%%%%%%%%%%%%%%%%%%%%%%%%%%%%%%%%%%%%%%%%%%%%%%%%%%%%%%%%%%%%%%%%%%%%%%%%%%%%%%%%%%%%%%%%%%%%%%%%%%%%
\subsection{Eavesdropping Models}
As depicted in Fig.~\ref{f1}, Eve acts as an external adversary that is assumed to have unlimited computational and quantum processing capabilities, while remaining subject to the fundamental laws of quantum mechanics. Bob’s receiver noise and internal imperfections are assumed to be trusted and inaccessible to Eve. On the other hand, Eve is assumed to perform collective Gaussian attacks against Gaussian-modulated CV-QKD protocols, which can be equivalently modeled as an entangling-cloner attack. In this model, Eve prepares two-mode squeezed vacuum (TMSV) states, also known as Einstein-Podolsky-Rosen (EPR) pairs, consisting of an injected mode and a retained idler mode\cite{weedbrook2012gaussian}.
The covariance matrix of a TMSV state is given by the following expression~\cite{ottaviani2020terahertz}:
\beq
\mathbf{\Sigma}_{\mathrm{EPR}} \triangleq
\begin{bmatrix}
    V_{e}\mathbf{I}_2 & \sqrt{V_{e}^2-1}\mathbf{Z}\\
     \sqrt{V_{e}^2-1}\mathbf{Z} & V_{e} \mathbf{I}_2
\end{bmatrix},
\label{eq12}
\eeq
where $\textbf{Z}$ is the Pauli-z matrix given as $\text{diag}\left(1,-1  \right)$ and $V_{e}$ denotes the variance of Eve’s EPR modes.

In an entangling-cloner attack, Eve injects one mode of a TMSV state into the environmental input of a lossy channel segment and stores the corresponding idler mode in her quantum memory. The complementary environmental output mode then constitutes Eve's observation of the signal leaked through that segment. In the considered RIS-assisted THz MIMO CV-QKD system, the relevant lossy propagation segments are the direct Alice-Bob path, the Alice-RIS path, and the RIS-Bob path, whose environmental-mode groups are denoted by $e_{{out}_j}$, $j\in\{d,t,r\}$ respectively.
%%%%%%%%%%%%%%%%%%%%%%%%%%%%%%%%%%%%%%%%%%%%%%%%%%%%%%%%%%%%%%%%%%
\subsubsection{Access-Constrained Segment-Wise Eavesdropping}

In the proposed access-constrained model, Eve is not assumed to voluntarily restrict her attack to a single propagation segment. Instead, Eve is assumed to collect all environmental modes that are physically accessible from her feasible deployment region. This assumption is practically valid for RIS-assisted THz CV-QKD because the direct Alice-Bob, Alice-RIS, and RIS-Bob links occupy different spatial propagation segments. In realistic deployments, these segments may not be simultaneously observable from a single feasible Eve location due to controlled regions around Alice, Bob, and the RIS, physical blockages, restricted-access areas, and the highly directional, short-range nature of THz propagation. Therefore, Eve's access constraint is not a limitation imposed on her signal processing capability; rather, it is a consequence of the physical deployment geometry and the segments that can be observed or probed from her accessible region.

Accordingly, Eve's side information is characterized by an access set $\mathcal{A}_E \subseteq \{d,t,r\}$, where $d$, $t$, and $r$ denote the direct Alice-Bob, Alice-RIS, and RIS-Bob propagation segments, respectively. Let $\mathcal{X}_E$ denotes Eve's feasible deployment region and $\mathcal{X}_j^{\rm obs}$ denotes the region from which the environmental modes of segment $j\in\{d,t,r\}$ can be physically observed, then the access set induced by an Eve location $\mathbf{x}_E$ can be written as
\beq
\mathcal{A}_E(\mathbf{x}_E)
=
\left\{j\in\{d,t,r\}: \mathbf{x}_E\in \mathcal{X}_j^{\rm obs} \right\}.
\label{eq13}
\eeq
Thus, if Eve can physically observe multiple segments, the corresponding multi-segment access set is included in the security analysis. The following segment-wise access cases are considered:
\begin{itemize}
\item \textbf{Single-segment access:} Eve accesses one environmental-mode
group, i.e., \(\mathcal{A}_E\in\{\{d\},\{t\},\{r\}\}\).
\item \textbf{Pairwise-segment access:} Eve simultaneously accesses two
environmental-mode groups, i.e.,
\(\mathcal{A}_E\in\{\{d,t\},\{d,r\},\{t,r\}\}\).
\item \textbf{Full segment access:} Eve accesses all three segment-wise
environmental-mode groups, i.e., \(\mathcal{A}_E=\{d,t,r\}\).
\end{itemize}

For a given access set \(\mathcal{A}_E\), Eve's quantum side information is formed by the environmental outputs and retained idler modes corresponding to all segments in \(\mathcal{A}_E\). The environmental modes associated with segments not included in \(\mathcal{A}_E\) still contribute to Bob's received noise through the physical channel model, but they are not available as Eve's quantum side information. Hence, the proposed model captures deployment-aware security scenarios in which Eve is computationally unrestricted across all accessible modes, while her quantum side information is determined by the propagation segments that are physically accessible from the network environment.
%%%%%%%%%%%%%%%%%%%%%%%%%%%%%%%%%%%%%%%%%%%%%%%%%%%%%%%%%%%%%%%%%%%
\subsubsection{Global Eavesdropping via Effective Channel Purification}
As a conservative benchmark, we also consider a global eavesdropping model in which Eve controls the purification of the effective end-to-end Alice-Bob channel. In this case, Eve is not restricted by segment-wise physical accessibility and is assumed to possess the environmental purification associated with the overall MIMO channel $\mathbf{H}$.
%%%%%%%%%%%%%%%%%%%%%%%%%%%%%%%%%%%%%%%%%%%%%%%%%%%%%%%%%%%%%%%%%%
%%%%%%%%%%%%%%%%%%%%%%%%%%%%%%%%%%%%%%%%%%%%%%%%%%%%%%%%%%%%%%%%%%
%%%%%%%%%%%%%%%%%%%%%%%%%%%%%%%%%%%%%%%%%%%%%%%%%%%%%%%%%%%%%%%%%%
%%%%%%%%%%%%%%%%%%%%%%%%%%%%%%%%%%%%%%%%%%%%%%%%%%%%%%%%%%%%%%%%%%
\section{Secret Key Rate Analysis}
The SKR in QKD quantifies the number of information-theoretically secure key bits that Alice and Bob can extract per channel use, after accounting for the information available to Eve under a considered attack model. In this work, the physical channel model remains fixed as described in Section~II, while different eavesdropping scenarios are distinguished by Eve's access to the segment-wise environmental modes.

We assume that Alice and Bob have access to the channel state information (CSI) required for precoding, combining, and SKR evaluation. This assumption allows us to isolate the impact of Eve's environmental-mode access on the achievable SKR \cite{sushilwcnc2026,11152330_sushil_RIS}. Let $\mathbf{H} = \mathbf{U}\mathbf{D}\mathbf{V}^\dagger$ be the singular value decomposition (SVD) of the overall channel $\mathbf{H}$.
Thus, with the intention of maximizing the data rate, Alice employs the precoder matrix as $\mathbf{V}$ before passing to the BS$_{a}$, and Bob employs its combiner as $\mathbf{U}^\dagger$ to one of the outputs of the BS$_{b}$.
Following the transmission of keys from Alice and the interaction with environmental modes, Bob receives the signal and performs measurements to decrypt the secret keys. In this context, Bob can perform two types of measurements, as follows: 1) homodyne, where he measures one of the two quadratures randomly, and 2) heterodyne, where he measures both quadratures simultaneously \cite{QC_book, collective_GA_Pirandola_2008}.
Since heterodyne detection introduces additional detector noise and previous studies have shown similar performance trends for the considered setting~\cite{11152330_sushil_RIS}, we focus on homodyne detection. Consequently, Bob's output vector is given as
\beqarr
&&\hspace{-0.71cm}\mathbf{b}=\mathbf{U}^\dagger\left(\sqrt{\eta_a\eta_b}\mathbf{H}_\text{d}-\sqrt{\left(1-\eta_a\right)\left(1-\eta_b\right)}\mathbf{H}_\text{r}\Phi\mathbf{H}_\text{t}\right)\mathbf{V}\mathbf{a}\nn \\
&&+\mathbf{U}^\dagger\left(\sqrt{\left(1-\eta_a\right) \eta_b}\mathbf{H}_\text{d}+\sqrt{\eta_a\left(1-\eta_b\right)} \mathbf{H}_\text{r}\Phi\mathbf{H}_\text{t}\right) \mathbf{v}_0\nn\\
&&+\sqrt{\eta_b}\mathbf{U}^\dagger\mathbf{N}_\text{d}\mathbf{e}_\text{d} +\sqrt{1-\eta_b}\mathbf{U}^\dagger \mathbf{H}_\text{r}\Phi \mathbf{N}_\text{t} \mathbf{e}_\text{t}\nn\\
&& +\sqrt{1-\eta_b} \mathbf{U}^\dagger \mathbf{N}_\text{r} \mathbf{e}_\text{r}+\mathbf{U}^\dagger \mathbf{n}_\textrm{b} \, .
\label{eq14}
\eeqarr
The three environmental terms in \eqref{eq14} correspond to the direct Alice-Bob, Alice-RIS, and RIS-Bob propagation segments, respectively. Hence, the physical received signal at Bob is common to all eavesdropping cases, whereas Eve's side information depends on the environmental-mode set accessible to her.
Following the homodyne measurement, a reconciliation technique is employed by Bob to fix errors, which are typically of two types, namely, 1) direct reconciliation (DR) and 2) RR \cite{2_ref_paper, ref-7}. RR works better than DR because the RR protocol can achieve positive SKR for any values of transmittance between $0$ and $1$, while DR needs more than $50\%$ of the transmittance value to achieve a positive SKR. Owing to this reason, we consider Bob to employ the RR protocol.
%%%%%%%%%%%%%%%%%%%%%%%%%%%%%%%%%%%%%%%%%%%%%%%%%%%%%%%%%%%%%%%%%%
%%%%%%%%%%%%%%%%%%%%%%%%%%%%%%%%%%%%%%%%%%%%%%%%%%%%%%%%%%%%%%%%%%
\subsection{Access-Constrained Secret Key Rate}
Taking into account the homodyne measurement and RR at Bob's end, together with the eavesdropping models of Section~II.C, the SKR for a given access set is
\beqarr
&&\hspace{-1cm}\textrm{SKR}_{\mathcal{A}_E} = \beta_{\rm rec}\, \mathcal{I}(A;B) - \chi\left(B;E_{\mathcal{A}_E}\right) \, , \nn \\
&& \hspace{-1cm} \mathcal{A}_E\in \left\{
\{d\},\{t\},\{r\},
\{d,t\},\{d,r\},\{t,r\},
\{d,t,r\}
\right\},
\label{eq15}
\eeqarr
where $\beta_{rec}$ is the reconciliation efficiency and $E_{\mathcal{A}_E}$ collects all environmental output and retained idler modes of the segments in $\mathcal{A}_E$, arising from the deployment geometry \eqref{eq13}.

Further, in \eqref{eq14}, $\mathcal{I}(A;B)$ denotes the classical mutual information between Alice and Bob, and $\chi\left(B;E_{\mathcal{A}_E}\right)$ represents the Holevo (quantum) information of Bob’s and Eve’s quantum states.
The achievable mutual information between Alice and Bob is given as 
\beqarr
\mathcal {I}\left(A;B\right) = \mathcal {H}\left(B\right) - \mathcal {H}\left(B|A\right), 
\label{eq16}
\eeqarr
where $\mathcal {H}\left(B\right)$ is the Shannon entropy, expressed as
\beq
\mathcal{H} \left(B\right) = - \int_{\mathbb{C}^{N_R}}
f \left( \textbf{b} \right)
\log_2 \left( f \left( \textbf{b} \right) \right) \, ,
\label{eq17}
\eeq
and $\mathcal {H}\left(B|A\right)$ is the conditional Shannon entropy computed as
\beq
\mathcal{H} \left(B|A\right) = - \int_{\mathbb{C}^{N_R}}
f \left( \mathbf{b} \big| \mathbf{a}  \right)
\log_2 \left( f \left( \mathbf{b} \big| \mathbf{a} \right) \right) \, .
\label{eq18}
\eeq
Here, $f(\mathbf{b})$ and $f(\mathbf{b} | \mathbf{a})$ represent the probability density functions (p.d.f.s) of the vector $\textbf{b}$ and the conditional p.d.f. of $\textbf{b}$ given $\mathbf{a}$, respectively, which can be computed from (\ref{eq14}).
It is important to note that calculating these p.d.f.s requires the corresponding covariance matrices $\mathbf{\Sigma}_{\textbf{b}}$ and $\mathbf{\Sigma}_{\textbf{b}|\textbf{a}}$, which are obtained as 
\begin{align}
\mathbf{\Sigma}_{\textbf{b}}
& = 
V_a \mathbf{U}^\dagger\left(\sqrt{\eta_a\eta_b}\mathbf{H}_\text{d}-\sqrt{\left(1-\eta_a\right)\left(1-\eta_b\right)} \mathbf{H}_\text{r} \Phi\mathbf{H}_\text{t}\right) 
\nn\\ 
& \times \left(\sqrt{\eta_a\eta_b}\mathbf{H}_\text{d}-\sqrt{\left(1-\eta_a\right)\left(1-\eta_b\right)} \mathbf{H}_\text{r} \Phi\mathbf{H}_\text{t}\right)^\dagger\mathbf{U}
\nn\\ &
+V_{v_0} \left(\sqrt{\left(1-\eta_a\right)\eta_b}\mathbf{H}_\text{d}
+\sqrt{\eta_a\left(1-\eta_b\right)} \mathbf{H}_\text{r} \Phi\mathbf{H}_\text{t}\right) 
\nn\\ 
& \times \left(\sqrt{\left(1-\eta_a\right)\eta_b}\mathbf{H}_\text{d}
+\sqrt{\eta_a\left(1-\eta_b\right)} \mathbf{H}_\text{r} \Phi\mathbf{H}_\text{t}\right)^\dagger
\nn\\ 
& + V_{e_d} \eta_b \mathbf{U}^\dagger \mathbf{N}_\text{d}\mathbf{N}_\text{d}^\dagger \mathbf{U}
+V_{e_r} \left(1-\eta_b\right) \mathbf{U}^\dagger \mathbf{N}_\text{r}\mathbf{N}_\text{r}^\dagger \mathbf{U} \nn \\
& +V_{e_t} \left(1-\eta_b\right) \mathbf{U}^\dagger \mathbf{H}_{r} \Phi \mathbf{N}_\text{t}\mathbf{N}_\text{t}^\dagger \Phi^\dagger \mathbf{H}_{r}^\dagger \mathbf{U}
+ \sigma_{\mathrm b}^2 \mathbf{I}_{N},
\label{eq19}
\end{align}
and
\begin{align}
\mathbf{\Sigma}_{\mathbf{b}|\mathbf{a}} &=
V_0 \mathbf{U}^\dagger\left(\sqrt{\eta_a\eta_b}\mathbf{H}_\text{d}-\sqrt{\left(1-\eta_a\right)\left(1-\eta_b\right)} \mathbf{H}_\text{r} \Phi\mathbf{H}_\text{t}\right) 
\nn\\ 
& \times \left(\sqrt{\eta_a\eta_b}\mathbf{H}_\text{d}-\sqrt{\left(1-\eta_a\right)\left(1-\eta_b\right)} \mathbf{H}_\text{r} \Phi\mathbf{H}_\text{t}\right)^\dagger\mathbf{U}
\nn\\ &
+V_{v_0} \left(\sqrt{\left(1-\eta_a\right)\eta_b}\mathbf{H}_\text{d}
+\sqrt{\eta_a\left(1-\eta_b\right)} \mathbf{H}_\text{r} \Phi\mathbf{H}_\text{t}\right) 
\nn\\ 
& \times \left(\sqrt{\left(1-\eta_a\right)\eta_b}\mathbf{H}_\text{d}
+\sqrt{\eta_a\left(1-\eta_b\right)} \mathbf{H}_\text{r} \Phi\mathbf{H}_\text{t}\right)^\dagger
\nn\\ 
& + V_{e_d} \eta_b \mathbf{U}^\dagger \mathbf{N}_\text{d}\mathbf{N}_\text{d}^\dagger \mathbf{U}
+V_{e_r} \left(1-\eta_b\right) \mathbf{U}^\dagger \mathbf{N}_\text{r}\mathbf{N}_\text{r}^\dagger \mathbf{U} \nn \\
& +V_{e_t} \left(1-\eta_b\right) \mathbf{U}^\dagger \mathbf{H}_{r} \Phi \mathbf{N}_\text{t}\mathbf{N}_\text{t}^\dagger \Phi^\dagger \mathbf{H}_{r}^\dagger \mathbf{U}
+ \sigma_{\mathrm b}^2 \mathbf{I}_{N},
\label{eq20}
\end{align}
where $N=\min\{N_T, N_R,K\}$, $V_a =\left(V_s + V_0 \right)$, $V_s$ is the variance of Alice signal state and $V_0$ represents the variance of the preparation vacuum state, $V_{v_0}$ denotes the variance of the vacuum state $v_0$ at BS$_{a}$, and $V_{e_j}$ is the variance of the environmental mode associated with segment $j$. Using (\ref{eq16})-(\ref{eq20}) followed by algebraic simplifications, the mutual information in (\ref{eq16}) is obtained as 
\beq 
\mathcal {I}\left(A;B\right)= \frac{1}{2}\log _{2} \Bigg | \frac{\mathbf{\Sigma}_{\mathbf{b}}}{\mathbf{\Sigma}_{\mathbf{b}|\mathbf{a}}} \Bigg|.
\label{eq21}
\eeq
It is important to note that $\mathcal I(A;B)$ is evaluated from Bob's complete received signal and is therefore common to all Eve access cases. The distinction among the different eavesdropping scenarios appears in the Holevo information term $\chi\left(B;E_{\mathcal{A}_E}\right)$.
%%%%%%%%%%%%%%%%%%%%%%%
\begin{figure}[!t]
    \centering
    \includegraphics[width=9cm,height=5.5cm]{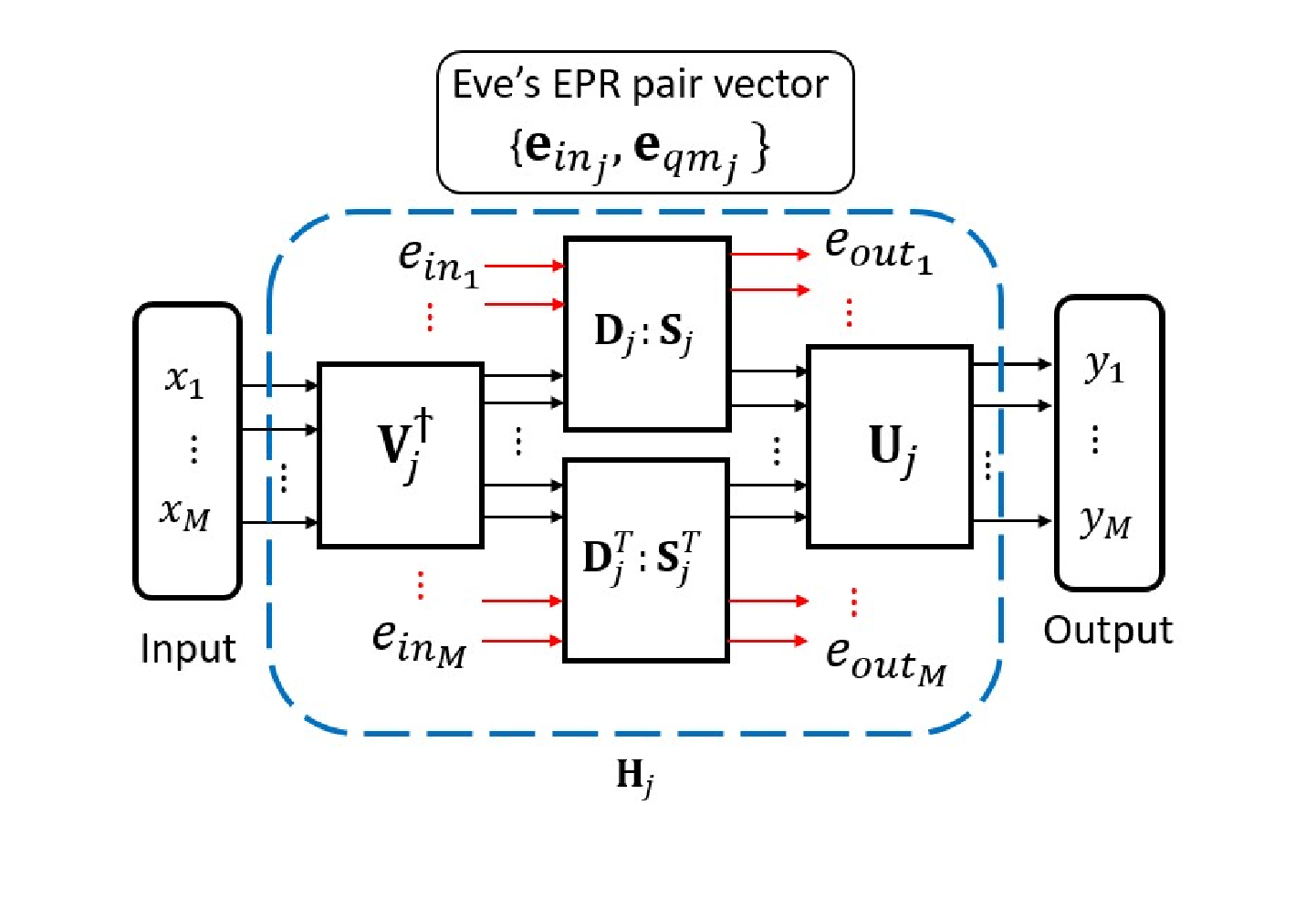} \vspace{-1cm}
    \caption{A depiction of the four beam-splitter dilation used to model the segment-wise loss of the considered MIMO channel}
    \label{f2}
\end{figure}
%%%%%%%%%%%%%%%%%%%%%%%%%%%%%%
To evaluate $\chi\left(B;E_{\mathcal{A}_E}\right)$ for arbitrary access-constrained eavesdropping with $\mathcal{A}_E \in \{\{d\},\allowbreak
\{t\},\allowbreak
\{r\},\allowbreak
\{d,t\},\allowbreak
\{d,r\},\allowbreak
\{t,r\},\allowbreak
\{d,t,r\}\}$, we first characterize the environmental output modes available to Eve by modeling a lossy channel with the beam splitters as depicted in Fig.~\ref{f2}, these modes are given as
\beqarr
&&\hspace{-0.61cm}\mathbf{e}_{\mathrm{out},j} = \nn \\
&& \hspace{-0.68cm}\begin{cases}
-\sqrt{\eta_a}\mathbf{S}_{d} \mathbf{V}_{d}^\dagger \mathbf{V}\mathbf{a}
+\sqrt{\left(1-\eta_a\right)}\mathbf{S}_{d} \mathbf{V}_{d}^\dagger \mathbf{v}_0
+ \mathbf{D}_{d} \mathbf{e}_{in_d}  & \!\!\!\!, j= d,
\\
-\sqrt{\left(1-\eta_a\right)}\mathbf{S}_{t} \mathbf{V}_{t}^\dagger \mathbf{V}\mathbf{a}
+\sqrt{\eta_a}\mathbf{S}_{t} \mathbf{V}_{t}^\dagger\mathbf{v}_0
+ \mathbf{D}_{t} \mathbf{e}_{in_t} & \!\!\!\! , j= t,
\\
-\sqrt{\left(1-\eta_a\right)}\mathbf{S}_{r}\mathbf{V}_{r}^\dagger\Phi \mathbf{H}_{t} \mathbf{V} \mathbf{a} 
+\sqrt{\eta_a}\mathbf{S}_{r}\mathbf{V}_{r}^\dagger\Phi \mathbf{H}_{t} \mathbf{v}_0\\
-\mathbf{S}_{r}\mathbf{V}_{r}^\dagger \Phi\mathbf{U}_{t}\mathbf{S}_{t} \mathbf{e}_{{in}_t}
+ \mathbf{D}_{r} \mathbf{e}_{in_r} &\hspace{-0.4cm}, \ j = r \, ,
\end{cases}\nn\\
\label{eq22}
\eeqarr
where matrix $\textbf{D}_j$ and $\textbf{S}_j$ are given as
\bsub
\beq
\mathbf{D}_{j} =
\begin{bmatrix}
    \text{diag}\left(\sqrt{\beta_{j,1}},\ldots,\sqrt{\beta_{j,{r_H}}}\right) & \textbf{0}_{{r_H} \times \left({N_j}-r_H\right)}\\
    \textbf{0}_{\left({N_j}-r_H\right) \times {r_H} } & \textbf{0}_{\left({N_j}-r_H\right) \times \left({N_j}-r_H\right)}
\end{bmatrix},
\label{eq23a}
\eeq
and
\beq
\mathbf{S}_j =
\text{diag}\left(\sqrt{1-\beta_{j,1}}, \ldots, \sqrt{1-\beta_{j,{r_H}}}, \underbrace{1,\ldots,1}_{\left(N_j-{r_H}\right) \text{ times}} \right) \, ,
\label{eq23b}
\eeq 
\esub
where $\sqrt{\beta_{j,1}},...,\sqrt{\beta_{j,{r_H}}}$ are the $r_H$ non-zero singular values of the $j$-th channel matrix. Consequently the environmental coupling matrices $\mathbf{N}_{j}$ for the corresponding channel are given as
\beq
\mathbf{N}_{j} = \mathbf{U}_{j}\mathbf{S}_{j} \, ,\qquad j\in \{ d ,t, r \}.
\label{eq24}
\eeq

The Holevo information between Bob and Eve for the access-constrained eavesdropping scenario in $\mathcal{A}_E$ is given by
\beq
\chi\left(B;E_{\mathcal{A}_E}\right)
= \mathcal{S} \left(E_{\mathcal{A}_E}\right)
- \mathcal{S} \left(E_{\mathcal{A}_E}|B\right),
\label{eq25}
\eeq
where $\mathcal{A}_E \in \left\{ \{d\},\{t\},\{r\}, \{d,t\},\{d,r\},\{t,r\}, \{d,t,r\} \right\}$, $\mathcal{S} \left(E_{\mathcal{A}_E}\right)$  and $\mathcal{S} \left(E_{\mathcal{A}_E}|B\right)$ represent the von Neumann (quantum) entropy of Eve's access modes and the conditional von Neumann entropy of Eve's access modes given Bob's received modes. The von Neumann entropy is computed as 
\beq
\mathcal{S} \left(E_{\mathcal{A}_E}\right) \text{ (or) }\mathcal{S} \left(E_{\mathcal{A}_E}|B\right)  =  \sum_{i=1}^{N_{\mathcal{A}_E}} h_o(\lambda_{i}) \, ,
\label{eq26}
\eeq
where $\lambda_{i}$s $\geq1$ are the symplectic eigenvalues of the correlation matrices $\Sigma_{E_{\mathcal{A}_E}}$ and $\Sigma_{E_{\mathcal{A}_E}|B}$ of $\mathcal{A}_E$, and the function $h_o (\cdot)$ is given as
\beqarr
\!\!\!\!\!\!\!\! \!h_o (\lambda ) \! \! \! \! \! &=& \!\!\! \! \! \! \left(\frac{\lambda+1}{2}\right)
\!\log_2 \!\left(\! \frac{\lambda+1}{2}\! \right) \!\!-\!\! \left(\! \frac{\lambda-1}{2} \!\right)\!
\log_2 \!\left( \!\frac{\lambda-1}{2} \!\right)\!\!.
\label{eq27}
\eeqarr
%%%%%%%%%%%%%%%%%%%%%%%%%%%%%%%%%%%%%%%%%%%%%%%%%%%%%%%%%%%%%%%%%%%%%%%%%%%%%%%%%%%%%%%%%%%%%%%%%%%%%%%%%%%%%%%%%%%%%%%%%%%%%%%%%%%%%%%%%%%%%%%%%%%%%%%%%%%%%%%%%%%%%%%%%%%%%%%%%%%%%%%%%%%%%%%%%%%%%%%%%%%%%%%%%%%%%%%%%%%%%%
\subsubsection{Single-Segment Access}

For the single-segment access case $\mathcal{A}_E=\{j\}$, where $j \in \{d, t, r\}$, Eve's accessible quantum system is composed of the environmental output mode $\mathbf{e}_{\mathrm{out}, j}$ and the retained idler mode $\mathbf{e}_{\mathrm{qm}, j}$. The corresponding covariance matrix is given as
\beq
\mathbf{\Sigma}_{E_j}=
\begin{bmatrix}
   \mathbf{\Sigma}_{\textbf{e}_{\text{out}_{j}}} &  \mathbf{\Sigma}_{\textbf{e}_{\text{oq}_{j}}}\\
\mathbf{\Sigma}_{\textbf{e}_{\text{oq}_{j}}}^\dagger & \mathbf{\Sigma}_{\textbf{e}_{\textrm{qm}_{j}}}
\end{bmatrix}, j\in\{d,t,r\},
\label{eq28}
\eeq
where $\mathbf{\Sigma}_{\textbf{e}_{\text{qm}_{j}}}= V_{e_j} \mathbf{I}_{N_j}$, $\mathbf{\Sigma}_{\textbf{e}_{\text{oq}_j}}= \sqrt{V_{e_j}^2 -1} \, \mathbf{D}_{j}$, $j\in\{d,t,r\}$, and
\beqarr
&&\hspace{-0.61cm}\mathbf{\Sigma}_{\textbf{e}_{\text{out}_j}} = \nn \\
&& \hspace{-0.61cm}\begin{cases}
\eta_a V_a\mathbf{S}_{d} \mathbf{S}_{d}^\dagger
+\left(1-\eta_a\right)V_{v_0}\mathbf{S}_{d} \mathbf{S}_{d}^\dagger
+ V_{e_d}\mathbf{D}_{d}\mathbf{D}_{d}^\dagger  & \hspace{-0.6cm}, j= d,
\\
\left(1-\eta_a\right)V_a\mathbf{S}_{t} \mathbf{S}_{t}^\dagger 
+\eta_aV_{v_0}\mathbf{S}_{t} \mathbf{S}_{t}^\dagger
+ V_{e_t}\mathbf{D}_{t}\mathbf{D}_{t}^\dagger & \hspace{-0.6cm}, j= t,
\\
\left(\left(1-\eta_a\right)V_a +\eta_aV_{v_0} \right)
\left(\mathbf{S}_{r}\mathbf{V}_{r}^\dagger\Phi \mathbf{H}_{t}\right)\left(\mathbf{S}_{r}\mathbf{V}_{r}^\dagger\Phi \mathbf{H}_{t}\right)^\dagger+
\\
V_{e_t}\left(\mathbf{S}_{r}\mathbf{V}_{r}^\dagger \Phi\mathbf{U}_{t}\mathbf{S}_{t}\right)\left(\mathbf{S}_{r}\mathbf{V}_{r}^\dagger \Phi\mathbf{U}_{t}\mathbf{S}_{t}\right)^\dagger+ V_{e_r}\mathbf{D}_{r} \mathbf{D}_{r}^\dagger&\hspace{-0.6cm},j = r \, .
\end{cases}\nn\\
\label{eq29}
\eeqarr

Furthermore, Eve's conditional von Neumann entropy $\mathcal {S}\left(E_{j}|B\right)$ is obtained from the correlation matrix generated with the $\mathbf{b}$, $\mathbf{e}_{\text{out}_j}$, and $\mathbf{e}_{\text{qm}_j}$ is given as
\beq
\mathbf{\Sigma}_{E_jB}^{\text{joint}}=
\begin{bmatrix}
    \mathbf{\Sigma}_{E_j} & \mathbf{\Sigma}_{E_j B} \\ 
    \mathbf{\Sigma}_{E_jB}^\dagger  & \mathbf{\Sigma}_{\mathbf{b}} 
\end{bmatrix}\, , j\in\{d,t,r\}.
\label{eq30}
\eeq
This leads to the conditional covariance matrix of Eve's modes that are dependent on the outcomes of Bob’s homodyne measurements, which is represented as
\beq
\mathbf{\Sigma}_{E_j|B} = \mathbf{\Sigma}_{E_j} - \mathbf{\Sigma}_{E_j B} \left(\Pi\left(\mathbf{\Sigma}_{\textbf{b}}\otimes \mathbf{I}_2\right)\Pi\right)^{+} \mathbf{\Sigma}_{E_jB}^\dagger ,
\label{eq31}
\eeq
where $\Pi = \mathbf{I}_{N_j} \otimes 
\begin{pmatrix}
1 & 0 \\
0 & 0
\end{pmatrix}$, and $\mathbf{\Sigma}_{\mathbf{b}}$ and $\mathbf{\Sigma}_{E_j}$ are given in (\ref{eq19}) and (\ref{eq28}), respectively. Furthermore, $\mathbf{\Sigma}_{E_j B}$ is the quantum correlation of Eve's received mode ${\mathbf{e}_{\text{out}_j}}$ and Eve's stored mode in quantum memory ${\textbf{e}_{\text{qm}_j}}$ with Bob's output modes $\mathbf{b}$ is given as
\beq
\mathbf{\Sigma}_{E_jB}=
\begin{bmatrix}
     \mathbf{\Sigma}_{e_{o_j}b} \\
    \mathbf{\Sigma}_{e_{q_j}b}  
\end{bmatrix}
\, , j\in\{d,t,r\}\, ,
\label{eq32}
\eeq
where $\mathbf{\Sigma}_{e_{o_j}b}$ is given in (\ref{eq33}) on the top of the next page, and
\begin{figure*}
\beq
\mathbf{\Sigma}_{e_{o_j}b} = 
\begin{cases}
-\sqrt{\eta_b}\left(\eta_aV_a + \left(1-\eta_a\right)V_{v_0}\right)\mathbf{S}_{d} \mathbf{V}_{d}^\dagger \mathbf{H}_{d}^\dagger \mathbf{U}  + \sqrt{\eta_b} V_{e_d}\mathbf{D}_{d} \mathbf{S}_{d}^\dagger \mathbf{U}_{d}^\dagger \mathbf{U}\\
-\sqrt{\eta_a\left(1-\eta_a\right)\left(1-\eta_b\right)}\left(V_a-V_{v_0} \right)\mathbf{S}_{d} \mathbf{V}_{d}^\dagger \left( \mathbf{H}_{r} \Phi \mathbf{H}_{t} \right)^\dagger\mathbf{U}
& , j= d,
\\
\sqrt{\eta_a \eta_b\left(1-\eta_a\right)}\left(V_{v_0}-V_a\right) \mathbf{S}_{t} \mathbf{V}_{t}^\dagger \mathbf{H}_{d}^\dagger \mathbf{U} + \sqrt{\left(1-\eta_b\right)} V_{e_t}\mathbf{D}_{t} \mathbf{S}_{t}^\dagger \mathbf{U}_{t}^\dagger \Phi^\dagger \mathbf{H}_{r}^\dagger \mathbf{U}\\
+\sqrt{\left(1-\eta_b\right)}\left(\left(1-\eta_a\right)V_a+\eta_a V_{v_0} \right) \mathbf{S}_{t} \mathbf{V}_{t}^\dagger \left( \mathbf{H}_{r} \Phi \mathbf{H}_{t} \right)^\dagger \mathbf{U}
 & , j= t,
\\
\sqrt{\eta_a \eta_b\left(1-\eta_a\right)}\left(V_{v_0}-V_a\right) \mathbf{S}_{r} \mathbf{V}_{r}^\dagger \Phi \mathbf{H}_{t} \mathbf{H}_{d}^\dagger \mathbf{U} -\sqrt{\left(1-\eta_b\right)} V_{e_t}\mathbf{S}_{r} \mathbf{V}_{r}^\dagger \Phi \mathbf{U}_{t} \mathbf{S}_{t} \mathbf{S}_{t}^\dagger \mathbf{U}_{t}^\dagger \Phi^\dagger \mathbf{H}_{r}^\dagger\mathbf{U}\\
+\sqrt{\left(1-\eta_b\right)}\left(\left(1-\eta_a\right)V_a+\eta_a V_{v_0} \right) \mathbf{S}_{r} \mathbf{V}_{r}^\dagger \Phi \mathbf{H}_{t} \left( \mathbf{H}_{r} \Phi \mathbf{H}_{t} \right)^\dagger \mathbf{U} 
+ \sqrt{\left(1-\eta_b\right)} V_{e_r}\mathbf{D}_{r} \mathbf{S}_{r}^\dagger \mathbf{U}_{r}^\dagger\mathbf{U}
&,j = r
\end{cases}
\label{eq33}
\eeq
\noindent\rule{\textwidth}{.5pt}
\vspace{-0.8cm}
\end{figure*}
%%%%%%%%%%%%%%%%%%%%%%%%%
\beq
 \mathbf{\Sigma}_{e_{q_j}b} = 
\begin{cases}
\sqrt{\eta_b}\sqrt{V_{e_d}^2-1} \left(\mathbf{U}^\dagger \mathbf{U}_{d} \mathbf{S}_\text{d}\right)^{\dagger}  &, j = d , \\
\sqrt{\left(1-\eta_b\right)}\sqrt{V_{e_t}^2-1} \left(\mathbf{U}^\dagger\mathbf{H}_{r}\Phi \mathbf{U}_{t} \mathbf{S}_\text{t}\right)^{\dagger}  &,j = t,\\
\sqrt{\left(1-\eta_b\right)}\sqrt{V_{e_r}^2-1} \left(\mathbf{U}^\dagger \mathbf{U}_{r} \mathbf{S}_\text{r}\right)^{\dagger}  &, j = r .
\end{cases}
\label{eq34}
\eeq
Using (\ref{eq21}) and (\ref{eq26}), the SKR for RIS-assisted MIMO CV-QKD under the considered Eavesdropping model is given in (\ref{eq35}) on the top of the next page.
\begin{figure*}[!t]
\beqarr
\text{SKR}_{\mathcal{A}_E} \! \! \! \! &=& \! \! \! \!
\frac{\beta_{\rm rec}}{2}\log _{2} \Bigg | {\mathbf{I}}_{N} + V_s \mathbf{U}^\dagger \left(\sqrt{\eta_a\eta_b} \mathbf{H}_\text{d}-\sqrt{\left(1-\eta_a \right) \left(1-\eta_b\right)} \mathbf{H}_\text{r} \Phi\mathbf{H}_\text{t} \right)   \left(\sqrt{\eta_a\eta_b}\mathbf{H}_\text{d}-\sqrt{\left(1-\eta_a \right) \left(1-\eta_b\right)} \mathbf{H}_\text{r} \Phi \mathbf{H}_\text{t} \right)^\dagger \mathbf{U} 
\nn \\
&& \hspace{-1cm}\times \left(\mathbf{\Sigma}_{\mathbf{b}|\mathbf{a}}  \right)^{-1} \Bigg|- 
\left(\sum_{i=1}^{N_j} h_o\left(\lambda_{E_{\mathcal{A}_E},i}\right) -\sum_{i=1}^{N_j} h_o\left(\lambda_{E_{\mathcal{A}_E}|B,i}\right)\right), \, \mathcal{A}_E\in\{\{d\},\{t\}, \{r\}, \{d, t\}, \{d, r\}, \{t, r\}, \{d, t, r\}\}
\label{eq35}
\eeqarr
\noindent\rule{\textwidth}{.5pt}
\vspace{-0.8cm}
\end{figure*}
%%%%%%%%%%%%%%%%%%%%%%%%%%%%%%%%%%%%%%%%%%%%%%%%%%%%%%%%%%%%%%%%%%
\subsubsection{Pairwise-Segment Access}

For a pairwise access set $\mathcal{A}_E=\{i,j\}$, where $i,j\in\{d,t,r\}$ and $i\neq j$, Eve's covariance matrix is given as
\beq
\mathbf{\Sigma}_{E_{\{i,j\}}} =
\begin{bmatrix}
\mathbf{\Sigma}_{E_i} & \mathbf{\Sigma}_{E_i, E_j} \\
\mathbf{\Sigma}_{E_i, E_j}^{\dagger} & \mathbf{\Sigma}_{E_j}
\end{bmatrix} ,
\label{eq36}
\eeq
where $\mathbf{\Sigma}_{E_i}$ and $\mathbf{\Sigma}_{E_j}$, $i,j\in\{d,t,r\}$ given in \eqref{eq28}, the cross-covariance blocks are given as
\begin{subequations}
\begin{align}
\mathbf{\Sigma}_{E_d, E_t}
&=
\begin{bmatrix}
\mathbf{\Sigma}_{e_{o_d},{e_{o_t}}} & \mathbf 0 \\
\mathbf 0 & \mathbf 0
\end{bmatrix},
\label{eq37a} \\
\mathbf{\Sigma}_{E_d, E_r}
&=
\begin{bmatrix}
\mathbf{\Sigma}_{e_{o_d},{e_{o_r}}} & \mathbf 0 \\
\mathbf 0 & \mathbf 0
\end{bmatrix},
\label{eq37b} \\
\mathbf{\Sigma}_{E_t, E_r}
&=
\begin{bmatrix}
\mathbf{\Sigma}_{e_{o_t},{e_{o_r}}} & \mathbf 0 \\
\mathbf{\Sigma}_{e_{o_r},{e_{q_t}}}^{\dagger} & \mathbf 0
\end{bmatrix},
\label{eq37c}
\end{align}
\end{subequations}
where 
\begin{subequations}
\begin{align}
\mathbf{\Sigma}_{e_{o_d}, {e_{o_t}}}
&=
\sqrt{\eta_a\left(1-\eta_a\right)}V_a
\mathbf{S}_d \mathbf{V}_d^\dagger \mathbf{V}_t \mathbf{S}_t^\dagger \nn\\
&\quad
+
\sqrt{\eta_a\left(1-\eta_a\right)}V_{v_0}
\mathbf{S}_d \mathbf{V}_d^\dagger
\mathbf{V}_t \mathbf{S}_t^\dagger,
\label{eq38a} \\
\mathbf{\Sigma}_{e_{o_d}, {e_{o_r}}}
&=
\sqrt{\eta_a\left(1-\eta_a\right)}V_a
\mathbf{S}_d\mathbf{V}_d^\dagger \mathbf{H}_t^\dagger \mathbf{\Phi}^\dagger
\mathbf{V}_r\mathbf{S}_r^\dagger \nn\\
&\quad
+
\sqrt{\eta_a\left(1-\eta_a\right)}V_{v_0}
\mathbf{S}_d\mathbf{V}_d^\dagger
\mathbf{H}_t^\dagger \mathbf{\Phi}^\dagger
\mathbf{V}_r\mathbf{S}_r^\dagger,
\label{eq38b} \\
\mathbf{\Sigma}_{e_{o_t}, e_{o_r}}
&=
\left(1-\eta_a\right)V_a
\mathbf{S}_t\mathbf{V}_t^\dagger \mathbf{H}_t^\dagger \mathbf{\Phi}^\dagger
\mathbf{V}_r\mathbf{S}_r^\dagger \nn\\
&\quad
+
\eta_a V_{v_0}
\mathbf{S}_t\mathbf{V}_t^\dagger
\mathbf{H}_t^\dagger \mathbf{\Phi}^\dagger
\mathbf{V}_r\mathbf{S}_r^\dagger \nn\\
&\quad
- V_{e_t}
\mathbf{D}_t\mathbf{S}_t^\dagger
\mathbf{U}_t^\dagger \mathbf{\Phi}^\dagger
\mathbf{V}_r\mathbf{S}_r^\dagger
\label{eq38c} \\
\mathbf{\Sigma}_{e_{o_r},{e_{q_t}}}
&=
-\sqrt{V_{e_t}^{2}-1}
\mathbf{S}_r\mathbf{V}_r^\dagger \mathbf{\Phi} \mathbf{U}_t\mathbf{S}_t .
\label{eq38d}
\end{align}
\end{subequations}
In case of pairwise-segment access, the Eve's conditional covariance matrix $\mathbf{\Sigma}_{E_{\{i,j\}}|B}$ has the similar form as \eqref{eq31}, where $\mathbf{\Sigma}_{E_{\{i,j\}}}$ given in \eqref{eq36} and 
\beq
\mathbf{\Sigma}_{E_{\{i,j\}}B} =
\begin{cases}
\begin{bmatrix}
\mathbf{\Sigma}_{E_{d}B}\\
\mathbf{\Sigma}_{E_{t}B}
\end{bmatrix},& \{i,j\}=\{d,t\} ,
\\
\begin{bmatrix}
\mathbf{\Sigma}_{E_{d}B}\\
\mathbf{\Sigma}_{E_{r}B}
\end{bmatrix},& \{i,j\}=\{d,r\} ,
\\
\begin{bmatrix}
\mathbf{\Sigma}_{E_{t}B}\\
\mathbf{\Sigma}_{E_{r}B}
\end{bmatrix}, & \{i,j\}=\{t,r\} ,
\end{cases}
\label{eq39}
\eeq
where $\mathbf{\Sigma}_{E_{j}B}$, $j\in \{d, t, r\}$ is given in \eqref{eq32}.
%%%%%%%%%%%%%%%%%%%%%%%%%%%%%%%%%%%%%%%%%%%%%%%%%%%%%%%%%%%%%%%%%%
\subsubsection{Full Segment Access}

For a full segment access $\mathcal{A}_E=\{d,t,r\}$, Eve's covariance matrix is given as
\beq
\mathbf{\Sigma}_{E_{\{d, t, r\}}} =
\begin{bmatrix}
\mathbf{\Sigma}_{E_d} & \mathbf{\Sigma}_{E_d, E_t} &  \mathbf{\Sigma}_{E_d, E_r}
\\
\mathbf{\Sigma}_{E_d, E_t}^{\dagger} & \mathbf{\Sigma}_{E_t} & \mathbf{\Sigma}_{E_t, E_r}
\\
\mathbf{\Sigma}_{E_d, E_r}^\dagger& \mathbf{\Sigma}_{E_t, E_r}^\dagger & \mathbf{\Sigma}_{E_r}
\end{bmatrix} ,
\label{eq40}
\eeq
where $\mathbf{\Sigma}_{E_j}$, $j\in \{d, t, r\}$ given in \eqref{eq28} and $\mathbf{\Sigma}_{E_d, E_t}$, $\mathbf{\Sigma}_{E_d, E_r}$, and $\mathbf{\Sigma}_{E_t, E_r}$ are given in \eqref{eq37a}, \eqref{eq37b} and \eqref{eq37c}, respectively.
Further, the Eve's conditional covariance matrix $\mathbf{\Sigma}_{E_{\{d, t, r\}}|B}$ has the similar form as \eqref{eq31}, where $\mathbf{\Sigma}_{E_{\{d, t, r\}}}$ given in \eqref{eq40} and 
\beq
\mathbf{\Sigma}_{E_{\{d, t, r\}}B} =
\begin{bmatrix}
\mathbf{\Sigma}_{E_{d}B}\\
\mathbf{\Sigma}_{E_{t}B} \\
\mathbf{\Sigma}_{E_{r}B}
\end{bmatrix} ,
\label{eq41}
\eeq
where $\mathbf{\Sigma}_{E_{j}B}$, $j\in \{d, t, r\}$ is given in \eqref{eq32}. The access-constrained rates quantify deployment-dependent Eve access to segment-wise environmental modes. For reference, we also evaluate a global-purification benchmark in which Eve is assumed to purify the entire effective Alice-Bob channel.
%%%%%%%%%%%%%%%%%%%%%%%%%%%%%%%%%%%%%%%%%%%%%%%%%%%%%%%%%%%%%%%%%%
\subsection{Global-Purification Benchmark}
In addition to the access-constrained eavesdropping scenarios considered in the previous subsections, we now examine a benchmark case in which Eve is assumed to have access to the purification of the entire effective channel between Alice and Bob, and to apply a collective Gaussian attack to the overall channel.

In this setting, Eve is assumed to have full knowledge of the overall channel matrix $\mathbf{H}$, including perfect CSI. By employing the SVD and following standard CV-QKD analysis, Alice applies the precoder $\mathbf{V}$, and Bob applies the combiner $\mathbf{U}^\dagger$, thereby transforming the system into $r_H$ equivalent SISO channels. Consequently, the input-output relations for the $i$-th parallel channel are given by
\beq
{b_i} = \sqrt{\beta_i}a_{i} + \sqrt{1-\beta_i} e_{in_i} + n_{\text{b}_i}\, ,
\label{eq42}
\eeq
and
\beq
e_{o_i} = - \sqrt{1-\beta_i} a_{i} +\sqrt{\beta_i}e_{in_i}\, ,
\label{eq43}
\eeq
where $\sqrt{\beta_i}\, \, \forall i =1,\ldots r_H$ are the singular values of $\mathbf{H}$, with $r_H=\mathrm{rank}(\mathbf{H})$. Furthermore, $e_{\mathrm{in},i}$ is one mode of Eve’s TMSV state with variance $V_e$, $e_{o,i}$ is the corresponding environmental output mode accessible to Eve, and $n_{\mathrm{b},i}$ represents trusted Gaussian noise at Bob.

Consequently, the expression for the SKR under the scenario where Eve employs a collective Gaussian entanglement attack on the overall channel, and considering RR, is given as 
\beq
\textrm{SKR}_i = \beta_{\rm rec}\mathcal{I}\left(A_i;B_i\right) - \chi\left(B_i;E_i\right),
\label{eq44}
\eeq
where $\mathcal{I}\left(A_i; B_i\right)$ is given by
\beqarr
\mathcal{I}(A_i;B_i)=\frac{1}{2}\log_2\!\left(1+
\frac{\beta_i V_s}
{\beta_i V_0 + (1-\beta_i)V_e + \sigma_{\mathrm{b}}^2}\right) \, ,
\label{eq45}
\eeqarr
where $V_0$ is the vacuum noise variance, $V_e$ is the variance of Eve's TMSV modes, and $\sigma_{\mathrm{b}}^2$ denotes the trusted receiver noise variance at Bob.

Similarly, the mutual quantum information $\chi\left(B_i;E_i\right)$ \cite{ref-7}, is expressed as
\beq
\chi\left({B_i};E_i\right) = S \left(E_i \right)
- S \left( E_i \left|{B_i} \right. \! \right),
\label{eq46}
\eeq
where the corresponding covariance matrix $\mathbf{\Sigma}_{E_{i}}$ for each $i$-th parallel channel is expressed as
\beq
\mathbf{\Sigma}_{E_{i}} =
\begin{bmatrix}
    V_{e_{o_{i}}}\textbf{I}_2 & V_{e_{o_i}e_{\text{qm}_i}}\textbf{Z}\\
    V_{e_{o_i}e_{\text{qm}_i}} \textbf{Z}^T & V_e \textbf{I}_2
\end{bmatrix},
\label{eq47}
\eeq
where $V_{e_{o_{i}}} = \left(1-\beta_i\right) V_a + \beta_i V_e$ and $V_{{e_{o_i}}e_{\text{qm}_i}} = \sqrt{\beta_i\left(V_e^2-1\right)}$. This results in the symplectic eigenvalues of the covariance matrix of Eve's ancillary modes being given as \cite{ref-7}
\beq
\lambda_{i_{1,2}} = \sqrt{\frac{1}{2}
\left( \nabla_{i} \pm \sqrt{ \nabla^2_{i}
-4 \text{det} \left( \mathbf{\Sigma}_{E_{i}} \right)} \right)},
\label{eq48}
\eeq
where
\bsub
\beq
\nabla_{i} = 
V_{e_{o_{i}}}^2 + V_e^2 - 2 \beta_{i} \left( {V_e}^2-1 \right),
\label{eq49a}
\eeq
and 
\beq
\text{det} \left(\mathbf{\Sigma}_{E_{i}} \right)
\! = \! 
\left(V_{e_{o_{i}}} V_e
- \beta_{i} \left({V_e}^2-1\right) \right)^2.
\label{eq49b}
\eeq
\esub
Similarly, the conditional covariance matrix of Eve's state given Bob's quadrature, $\mathbf{\Sigma}_{E_{i}|{B_i}}$, is defined as follows~\cite{ref-7}:
\beq
\mathbf{\Sigma}_{E_{i}|{B_i}} \dn \mathbf{\Sigma}_{E_{i}} - \frac{1}{V_{b_i}}\textbf{W}_{i}\textbf{M} \textbf{W}_{i}^\dagger,
\label{eq50}
\eeq
where $\mathbf{\Sigma}_{E_{i}}$ is given in (\ref{eq47}) and
\beq
\textbf{M} =\begin{bmatrix}
    1&0\\
    0&0
\end{bmatrix}, \,
\textbf{W}_{i} = \begin{bmatrix}
     \langle E_{o_{i}} \cdot b_i \rangle \textbf{I}_2 \\
     \langle e_{qm_{i}}\cdot b_i\rangle \textbf{Z}
 \end{bmatrix}.
\label{eq51}
\eeq
From (\ref{eq46}) and (\ref{eq47}), we observe that $\mathbf{\Sigma}_{E_{i}|{B_i}}$ can be expressed in the form
\beq
\mathbf{\Sigma}_{E_{i}|{B_i}} =
\begin{bmatrix}
   \textbf{A}_{i} & \textbf{C}_{i} \\
    \textbf{C}^\dagger_{i} & \textbf{B}_{i}
\end{bmatrix},
\label{eq52}
\eeq
which is in a similar form as (\ref{eq47}) where
\beqarr
\textbf{A}_{i} \! \! \! \! &=& \! \! \! \!
\text{diag} \left( \frac{V_a V_e + \sigma_{\text{b}}^2 V_{e_{o_{i}}} }
{V_{b_i}},
V_{e{o_{i}}} \right), \nn \\
\textbf{B}_{i} \! \! \! \! &=& \! \! \! \!
\text{diag} \left( \frac{ \left(1-\beta_{i}
+ \left(\beta_{i} V_a + \sigma_{\text{b}}^2\right) V_e \right)}
{V_{b_i}}, V_e \right) , \nn \\
\textbf{C}_{i} \! \! \! \! &=& \! \! \! \!
\text{diag} \left( \frac{\left(V_a + \sigma_{\text{b}}^2 \right) V_{e_{o_i}e_{\text{qm}_i}}} {V_{b_i}},
- V_{e_{o_i}e_{\text{qm}_i}} \right).
\label{eq53}
\eeqarr
Thus, the symplectic eigenvalues of the conditional covariance matrix can be calculated as:
\beq
 \lambda_{i_{3,4}} =\sqrt{\frac{1}{2}\left( \tilde{\nabla}_{i}
\pm \sqrt{\tilde{\nabla}_{i}^2
- 4\text{det} \left(\mathbf{\Sigma}_{E_{i}|{B_i}} \right)}\right)},
\label{eq54}
\eeq
where
\bsub
\beqarr
\tilde{\nabla}_{i} \! \! \! \! &=& \! \! \! \!
\text{det} \left( \textbf{A}_{i} \right)
+ \text{det} \left( \textbf{B}_{i} \right)
- 2  \text{det} \left( \textbf{C}_{i} \right)  \, \nn \\
&=& \! \! \! \!
\frac{ \left( 1-\beta_{i} \right) V_e \left( V_a^2+1 \right)
+ 2 \beta_{i}V_a} 
{V_{b_i} } + \sigma_{\text{b}}^2 \nabla_i \, ,
\label{eq55a}
\eeqarr
and 
\beqarr
&& \! \! \! \! \! \! \! \! \!
\! \! \! \! \! \! \! \! \! \! 
\! \! \! \! \! \! \! \!
\text{det} \left(\mathbf{\Sigma}_{E_{i}|{B_{i}}} \right)
= \text{det} \left(\mathbf{\Sigma}_{E_{i}} \right) \nn \\
&& 
+ \frac{ \sigma_{\text{B}}^2 \Lambda(V_aV_e,1)  \left(\Xi + \Lambda(V_aV_e,1) \sigma_{\text{b}}^2  \right)} 
{V_{b_i}^2},
\label{eq55b}
\eeqarr
\esub
where $\Lambda(V_aV_e,1) = \left(1-\beta_{i} \right) V_a V_e + \beta_{i}$ and $\Xi =\Lambda(1,V_aV_e) V_{e_{o_{i}}} + V_a V_{e}^2 -2 V_a V_{e_{o_i}e_{\text{qm}_i}}^2 $.
Substituting all these results into (\ref{eq44}) and performing algebraic simplifications leads to the expression for the effective SKR of the RIS-assisted MIMO CV-QKD, where Eve performs a collective Gaussian attack on the overall channel, which is given in (\ref{eq56}) at the top of the next page.
\begin{figure*}[!t]
\beqarr
 \text{SKR}_{G}^{\text{MIMO}} = \!\sum_{i=1}^{r_H} \text{SKR}_{G, i}  \! \! \! \! &=& \! \! \! \! \!
 \sum_{i=1}^{r_H} \left( \frac{\beta_{\rm rec}}{2} \log_2 \left( 1 + \frac{ \beta_{i} V_s }
{\left( \beta_{i} V_0 + \left( 1 - \beta_{i} \right) V_e + \sigma_{\text{b}}^2 \right)} \right) 
 - h_o \left( \lambda_{1_{i}} \right)
- h_o \left( \lambda_{2_{i}} \right)
+ h_o \left( \lambda_{3_{i}} \right)
+ h_o \left( \lambda_{4_{i}} \right) \right) \nn \\
\label{eq56}
\eeqarr
\noindent\rule{\textwidth}{.5pt}
%\vspace{-0.8cm}
\end{figure*}
%%%%%%%%%%%%%%%%%%%%%%%%%%%%%%%%%%%%%%%%%%%%%%%%%%%%%%%%
\subsection{PSO-Based Joint RIS and Beam Splitters Optimization}
In alignment with the proposed RIS-assisted MIMO CV-QKD system and the analytically derived SKR expressions accounting for access-constrained eavesdropping and global-purification benchmark, the phase shifts of the RIS unit elements and values of $\eta_a$ and $\eta_b$ are optimized to enhance the system's secrecy performance. The goal of the optimization is to maximize the MIMO SKR by jointly adjusting the RIS phase shifts, $\eta_a$ and $\eta_b$, within their feasible ranges. Mathematically, the objective function is given as
\begin{align} \label{eq57}
        \mathcal{OP}:&\max_{\boldsymbol{\phi},\eta_a,\eta_b}
          \quad \text{SKR}_{\{\mathcal{A}_E, G\}}
          \\
        &\nonumber \, \text{\text{s}.\text{t}.}\, -\pi \leq \phi_k \leq \pi \, , \ \forall k \in \left\{1,\ldots,K \right\},\\
        &\nonumber \, \text{\text{s}.\text{t}.}\, 0 \leq \eta_{a} \leq 1 \,\ \text{and}\, \  0 \leq \eta_{b} \leq 1 \,.
\end{align}
Due to the highly nonlinear relationship between the SKR and the RIS phase matrix in the scenarios outlined in (\ref{eq35}) and (\ref{eq56}), a PSO-based algorithm (Algorithm 1) is used to solve the optimization problem presented in (\ref{eq57}). This approach aims to determine the optimal RIS phase configurations and the optimal $\eta_a$ and $\eta_b$, which are then used to evaluate the achievable SKR for each scenario considered.
%%%%%%%%%%%%%%%%%%%%%%%%%%%%%%%%%%%%%%%%%%%%%%%%%%%%%%%%%%%%%%%%%%%%%%%%%%%%%%%%%%%%%%%%%%%%%%%%%%%%%%%%%%%%%%%
\begin{algorithm}[!t]
\caption{PSO-Based $\boldsymbol{\Phi}$, $\eta_a$, and $\eta_b$ Joint Optimization} %for SKR Maximization (\ref{eq34}), (\ref{eq49})}
\begin{algorithmic}[1]
\State \textbf{Inputs:} $N_T$, $N_R$, number of RIS elements $K$, transmission distance, and PSO configuration parameters.
\State Specify the feasible search space for RIS phases as $-\pi \leq \phi_k \leq \pi$, $\forall k \in \{1,\ldots,K\}$, and $0\leq  \eta _a,\eta_b \leq 1$.
\For{each considered transmission distance}
    \State Generate the channel matrices $\mathbf{H}_{\mathrm{d}}$, $\mathbf{H}_{\text{t}}$, $\mathbf{H}_{\text{r}}$, and $\mathbf{H}$ using (\ref{eq2}).
    \State Define the fitness function as the SKR, $\text{SKR}_{\mathrm{\{\mathcal{A}_E, G\}}}$.
    \State Randomly initialize the particle swarm with random RIS phase configurations, $\eta_a$, and $\eta_b$.
    \For{each PSO iteration}
        \State Compute the fitness value $\text{SKR}_{\mathrm{\{\mathcal{A}_E, G\}}}$ for every particle.
        \State Update the personal-best and global-best solutions based on the maximum fitness achieved.
        \State Modify particle velocities and positions according to PSO update rules.
        \State Project updated phase, $\eta_a$, and $\eta_b$ values onto the feasible region defined in Step 2.
    \EndFor
    \State Obtain the optimized RIS phase matrix $\boldsymbol{\Phi}_{\mathrm{opt}}$, $\eta_a$, and $\eta_b$.
    \State Evaluate the corresponding $\text{SKR}_{\mathrm{\{\mathcal{A}_E, G\}}}^{\mathrm{max}}$.
\EndFor
\end{algorithmic}
\end{algorithm}
%%%%%%%%%%%%%%%%%%%%%%%%%%%%%%%%%%%%%%%%%%%%%%%%%%%%%%%%%%%%%%%%%%%%%%%%%%%%%%%%%%%%%%%%%%%%%%%%%%%%%%%%%%%%%%%%%%%%%%%%%%%%%%%%%%%%%%%%%%%%%%%%%%%%%%%%%%%%%%%%%%%%%%%%%%%%%%%%%%%%%%%%%%%%%%%%%%%%%%%%%%%%%%%%%%%%%%%%%%%%%%%%%%%%%%%%%%%%%%%%%%%%%%%%%%%%%%%%%%%%%%%%%%%%%%%%%%%%%%%%%%%%%%%%%%%%%%%%%%%%%%%%%%%%%%%%%%%%%%%%%%%%%%%%%%%%%%%%%%%%%%%%%%%%%%%%%%%%%%%%%%%%%%%%%%%%%%%%%%%%%%%%%%%%%%%%%%%%%%%%%%%%%%%%%%%%%%%%%%%%%%%%%%%%%%%%%%%%%%%%%%%%
\section{Numerical Results and Discussion}
This section presents the numerical results corroborating the analytical framework described and derived in the paper. For the simulation studies, we consider the standard system parameters as $\rho$ = 50 dB/Km, $T_e$ = 296 K (denoting the room temperature), antenna gain $G_a$ = 30 dBi, $\sigma_b^2=0.01$, reconciliation efficiency $\beta_{rec}=0.95$, the variance of Alice's initial modulated signal $V_s = 600$, the variance of the vacuum state $V_0=2\Bar{n}+1$ with $\Bar{n}=[\exp\left(hf_c/k_BT_e\right)-1]^{-1}$, where $h=6.626\times10^{-34} \text{J$\cdot$s}$ is the Planck's constant and $k_B=1.381 \times 10^{-23}$ $\text{JK}^{-1}$ is the Boltzmann's constant, the variance of Alice's quadrature $V_a = V_s+V_0$, $V_{v_0}=1$, and the variance of Eve's quadrature $V_{e_j} = 1$, $j\in\{d,t,r\}$ $\&$ $V_e = 1$.  
We also position the RIS at a distance of $0.3$ and $0.8$ times of the distance between Alice and Bob from transmitter and from receiver, respectively, to maximize its gain \cite{sushilwcnc2026} and consider two configurations of the phase shift matrix of the RIS: (a) $\boldsymbol{\Phi}_{\text{RIS, rand}}$ with all phases to be different and random; and (b) $\boldsymbol{\Phi}_{\text{RIS, opt}}$ as obtained using the PSO-based algorithm.
%%%%%%%%%%%%%%%%%%%%%%%%%%%%%%%%%%%%%%%%%%%%%%%%%%%%%%%%%%%%%%%
\begin{figure*}[!t]
     \centering
     \begin{subfigure}[b]{0.32\textwidth}
         \centering
        \includegraphics[width=\textwidth]{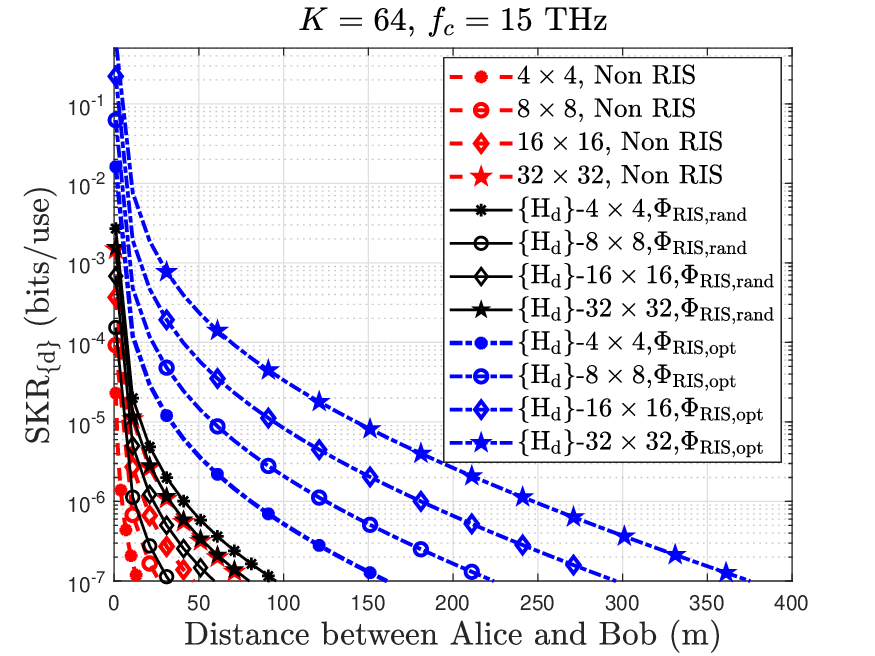}
         \caption{$\mathbf{H}_\text{d}$ segment, $\mathcal{A}_E=\{d\}$}
         \label{fig;3a}
     \end{subfigure}
     %\hspace{-0.3cm}
     \begin{subfigure}[b]{0.32\textwidth}
         \centering
        \includegraphics[width=\textwidth]{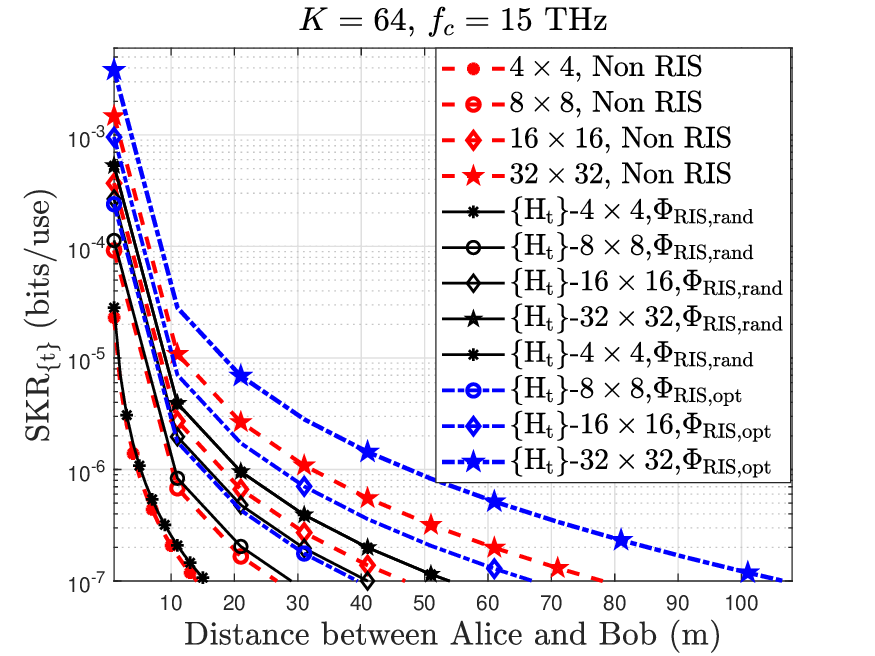}
         \caption{$\mathbf{H}_\text{t}$ segment, $\mathcal{A}_E=\{t\}$}
         \label{fig;3b}
     \end{subfigure}
     %\hspace{-0.3cm}
     \begin{subfigure}[b]{0.32\textwidth}
         \centering
        \includegraphics[width=\textwidth]{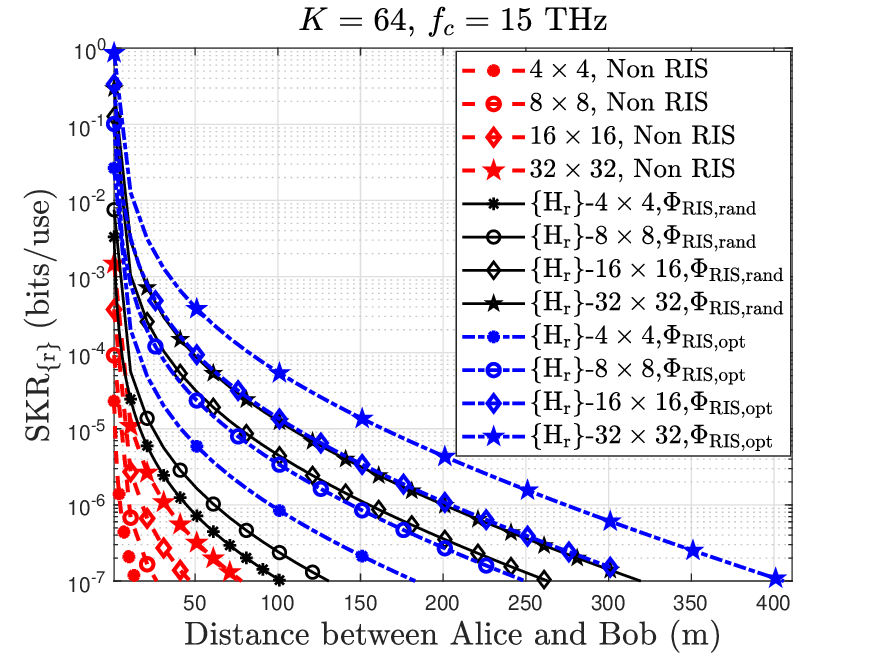}
         \caption{$\mathbf{H}_\text{r}$ segment, $\mathcal{A}_E=\{r\}$}
         \label{fig;3c}
     \end{subfigure}
     \centering
     \begin{subfigure}[b]{0.32\textwidth}
         \centering
        \includegraphics[width=\textwidth]{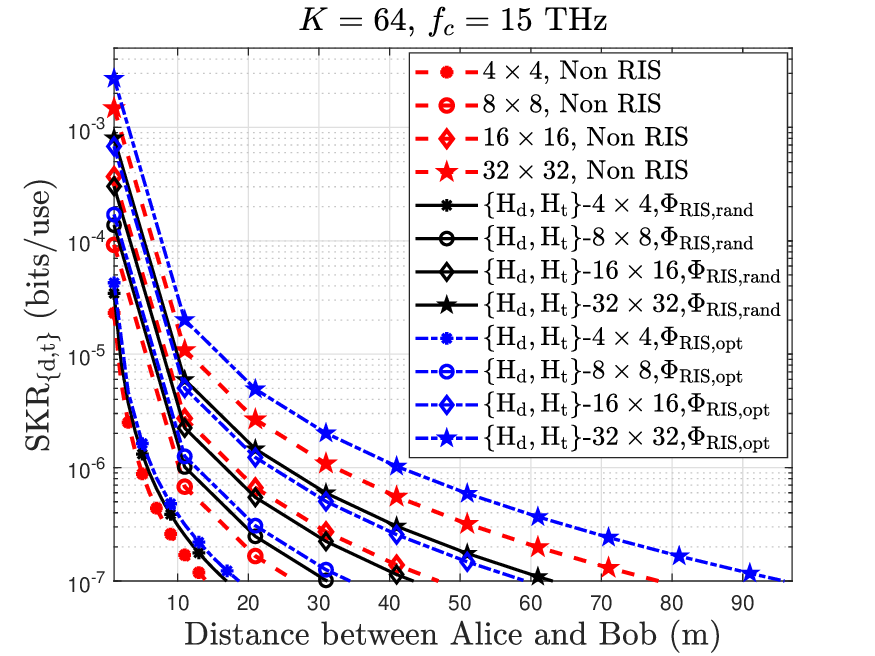}
         \caption{$\{\mathbf{H}_\text{d}, \mathbf{H}_\text{t}\}$ segment, $\mathcal{A}_E=\{d, t\}$}
         \label{fig;3d}
     \end{subfigure}
     %\hspace{-0.3cm}
     \begin{subfigure}[b]{0.32\textwidth}
         \centering
        \includegraphics[width=\textwidth]{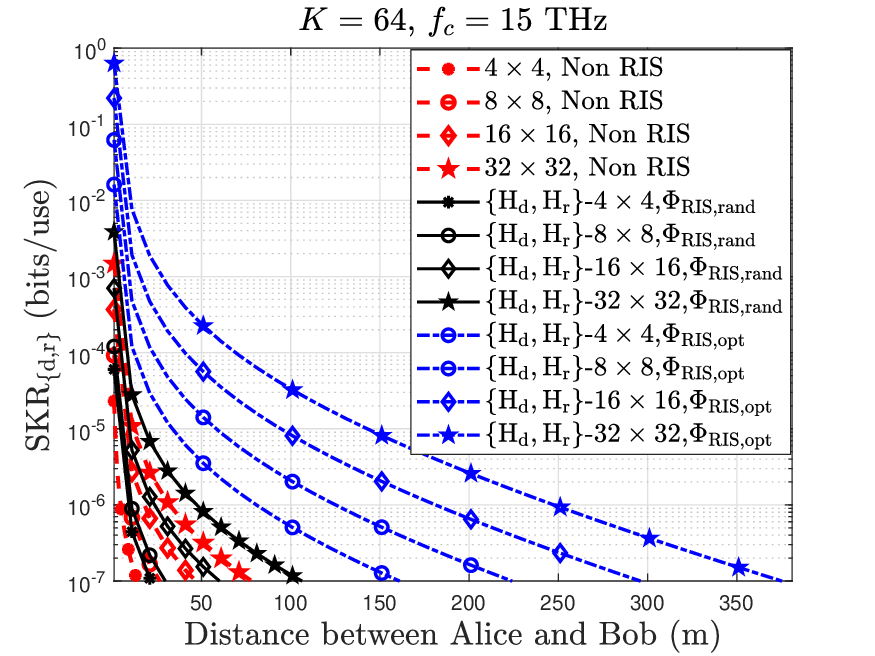}
         \caption{$\{\mathbf{H}_\text{d}, \mathbf{H}_\text{r}\}$ segment, $\mathcal{A}_E=\{d, r\}$}
         \label{fig;3e}
     \end{subfigure}
     %\hspace{-0.3cm}
     \begin{subfigure}[b]{0.32\textwidth}
         \centering
        \includegraphics[width=\textwidth]{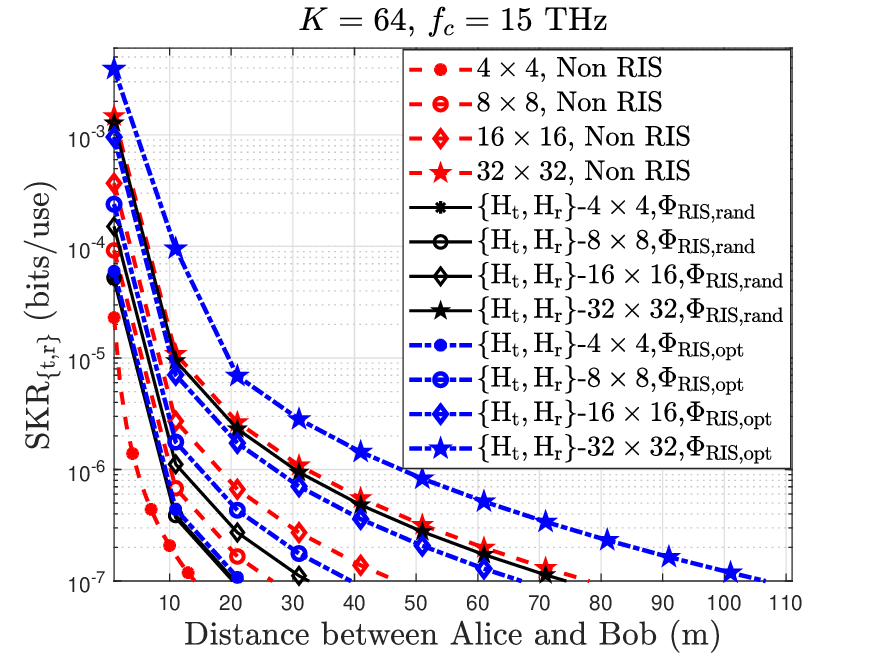}
         \caption{$\{\mathbf{H}_\text{t}, \mathbf{H}_\text{r}\}$ segment, $\mathcal{A}_E=\{t, r\}$}
         \label{fig;3f}
     \end{subfigure}
        \caption{SKR versus distance between Alice and Bob for $N_{R}=N_{T}=4, 8,16,32$, $K$ = 64, $f_c$ = 15 THz, and $d_a=0.5\lambda_c$ for Eve accesses environmental modes of (a) $\mathbf{H}_\text{d}$, (b) $\mathbf{H}_\text{t}$, (c) $\mathbf{H}_\text{r}$, (d) $\{\mathbf{H}_\text{d}, \mathbf{H}_\text{t}\}$, (e) $\{\mathbf{H}_\text{d}, \mathbf{H}_\text{r}\}$, and (f) $\{\mathbf{H}_\text{t}, \mathbf{H}_\text{r}\}$.}
        \label{f3}
        \vspace{-0.3cm}
\end{figure*}

Figure~\ref{f3} illustrates the variation in the SKR with the transmission distance between Alice and Bob across six access-constrained eavesdropping scenarios. Specifically, Figures~\ref{fig;3a}-\ref{fig;3c} correspond to the single-segment access cases $\mathcal{A}_E=\{\{d\},\{t\}, \{r\}\}$, respectively, whereas Figures~\ref{fig;3d}-\ref{fig;3f} present the pairwise access cases $\mathcal{A}_E=\{\{d,t\}, \{d,r\}, \{t,r\}\}$. It is observed that the SKR exhibits a similar performance trend across all six eavesdropping access sets. In these cases, the SKR decreases monotonically with increasing transmission distance due to path loss. However, a significant improvement in SKR is achieved as the MIMO configuration increases. This improvement can be attributed to the increased spatial degrees of freedom and array gain provided by larger MIMO configurations, which strengthen the legitimate channel and extend the achievable secure transmission range. Moreover, the optimized RIS configuration obtained using the proposed PSO algorithm consistently outperforms both the random RIS configuration and the conventional non-RIS system across all considered eavesdropping scenarios, demonstrating the effectiveness of jointly optimizing the RIS phase shifts. 

Among the single-segment access cases, the scenario $\mathcal{A}_E=\{t\}$ yields the lowest SKR, whereas $\mathcal{A}_E=\{d\}$ and $\mathcal{A}_E=\{r\}$ achieve comparatively higher secrecy rates. This observation indicates that the Alice-RIS propagation segment is the most security-sensitive part of the RIS-assisted transmission, since information leakage before reflection affects the overall effective channel more severely than leakage occurring on either the direct or the RIS-Bob link. When Eve simultaneously accesses two propagation segments, the SKR decreases further because additional correlated environmental modes become available for collective processing. However, the degradation is not determined solely by the number of compromised segments. Instead, it strongly depends on which propagation segments are jointly accessible. In particular, the access sets $\mathcal{A}_E=\{{d,t}\}$ and $\mathcal{A}_E=\{{t,r}\}$ exhibit noticeably lower SKRs than $\mathcal{A}_E=\{{d,r}\}$, confirming that the presence of the Alice-RIS segment dominates the secrecy degradation. In contrast, simultaneously accessing the direct and RIS-Bob segments without compromising the Alice-RIS link results in a relatively smaller performance loss. These results reveal that the transmitter-side RIS channel plays a dominant role in determining the secrecy performance of RIS-assisted THz MIMO CV-QKD systems.
%%%%%%%%%%%%%%%%%%%%%%%%%%%%%
\begin{figure}[!t]
    \centering
    \includegraphics[height=6cm,width=8cm]{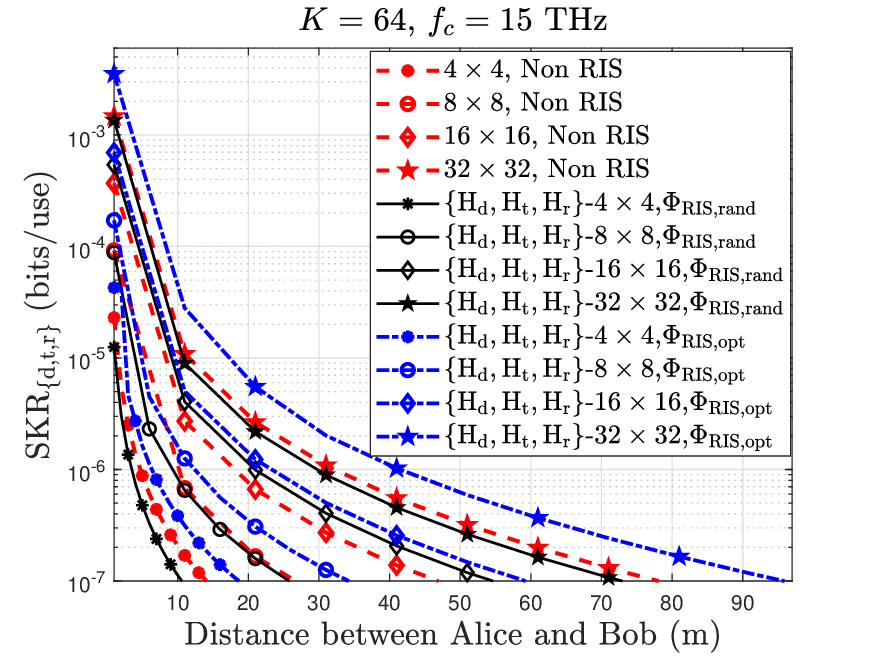}
    \caption{SKR versus distance between Alice and Bob for $N_{R}=N_{T}=\{4, 8, 16, 32\}$, $K$ = 64, $f_c$ = 15 THz, and $d_a=0.5\lambda_c$ for Eve accesses environmental modes of full-segment $\{\mathbf{H}_\text{d}, \mathbf{H}_\text{t}, \mathbf{H}_\text{r}\}$.}
    \label{f4}
\end{figure}

Figure~\ref{f4} illustrates the SKR when Eve simultaneously accesses the environmental modes of all three propagation segments, i.e., $\mathcal{A}_E=\{d,t,r\}$. This represents the strongest access-constrained attack considered in this paper and therefore yields the lowest SKR among all access-constrained eavesdropping scenarios. Despite Eve's increased side information, the proposed RIS-assisted system with PSO-based phase optimization consistently achieves higher SKR and longer secure transmission distances than the corresponding non-RIS system and the random-phase configuration across all considered MIMO configurations. Furthermore, increasing the antenna dimensions provides additional spatial diversity, leading to further improvements in the achievable SKR.
%%%%%%%%%%%%%%%%%%%%%%%%%%%%%%%%%%%%%%%
\begin{figure}[!t]
    \centering
    \includegraphics[height=6cm,width=8cm]{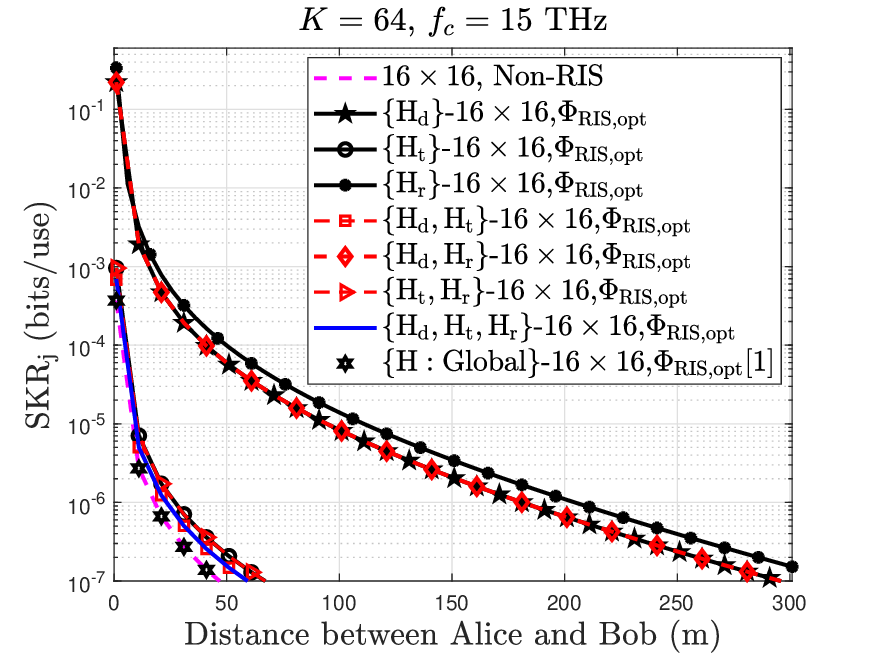}
    \caption{SKR versus distance between Alice and Bob for the non-RIS system, all seven access-constrained eavesdropping cases, and the global purification benchmark with $N_T=N_R=16$, $K=64$, and $f_c=15$~ THz.}
    \label{f5}
\end{figure}

Figure~\ref{f5} compares the SKR achieved under all seven access-constrained eavesdropping scenarios together with the conventional non-RIS system and the global eavesdropping benchmark. It is observed that the global eavesdropping model provides the lowest SKR, since Eve is assumed to have access to the purification of the entire effective channel. In contrast, all access-constrained scenarios achieve significantly higher SKRs because Eve is limited to environmental modes associated with physically accessible propagation segments. It is also observed that the access sets containing the Alice-RIS segment $\mathbf{H}_\text{t}$ exhibit lower SKRs than those without $\mathbf{H}_\text{t}$, indicating that compromising the transmitter-side RIS link leaks more useful information to Eve than compromising only the direct or RIS-Bob segments. Moreover, the optimized RIS phase configuration consistently outperforms the non-RIS system over the entire transmission range, demonstrating the security benefit of jointly optimizing the RIS under practical access-constrained eavesdropping.
%%%%%%%%%%%%%%%%%%%%%%%%%%%%%%%%%%%%%%%%%%%%%
\begin{figure*}
     \centering
     \begin{subfigure}[b]{0.33\textwidth}
         \centering
        \includegraphics[width=\textwidth]{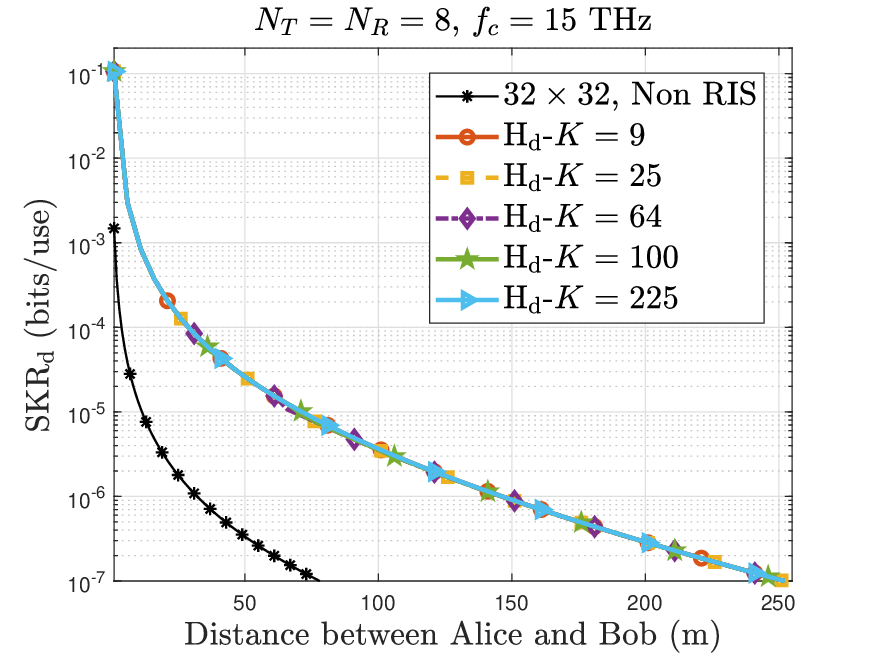}
         \caption{$\mathbf{H}_\text{d}$ segment, $\mathcal{A}_E=\{d\}$}
         \label{fig;6a}
     \end{subfigure}
     %\hspace{-0.3cm}
     \begin{subfigure}[b]{0.33\textwidth}
         \centering
        \includegraphics[width=\textwidth]{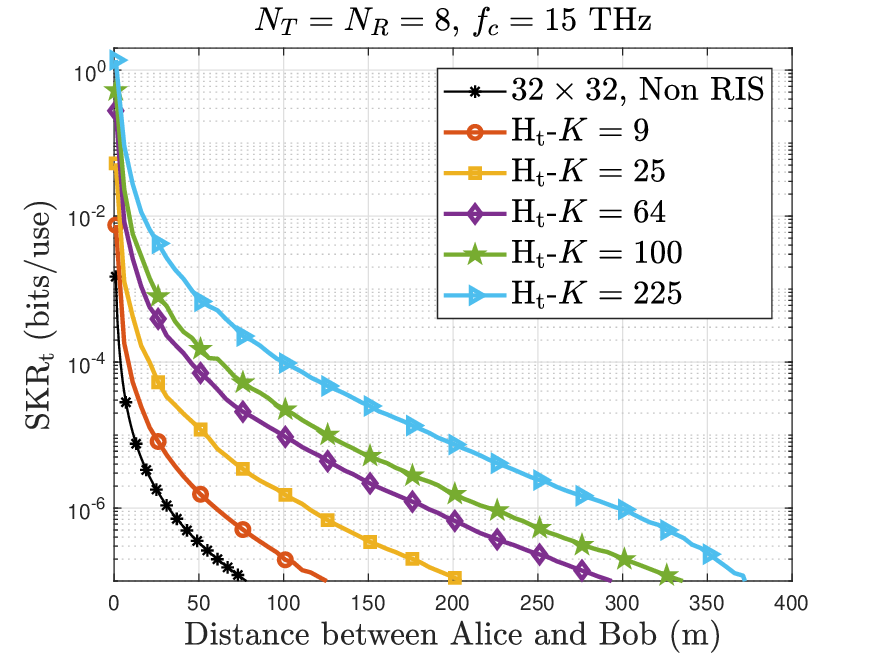}
         \caption{$\mathbf{H}_\text{t}$ segment, $\mathcal{A}_E=\{t\}$}
         \label{fig;6b}
     \end{subfigure}
     \hspace{-0.3cm}
     \begin{subfigure}[b]{0.33\textwidth}
         \centering
        \includegraphics[width=\textwidth]{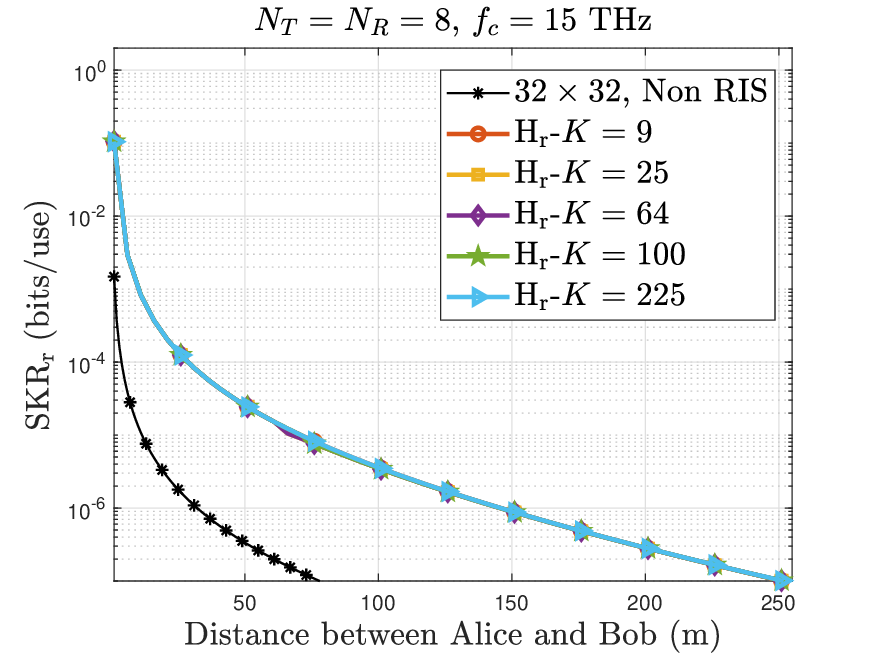}
         \caption{$\mathbf{H}_\text{r}$ segment, $\mathcal{A}_E=\{r\}$}
         \label{fig;6c}
     \end{subfigure}
        \caption{SKR versus distance between Alice and Bob for $K = 9,25, 64, 100, 225$ , $N_{R}=N_{T}=8$, $f_c = 15$ THz, and $d_a=0.5\lambda_c$ for Eve accesses environmental modes of (a) $\mathbf{H}_\text{d}$, (b) $\mathbf{H}_\text{t}$, and (c) $\mathbf{H}_\text{r}$.}
        \label{f6}
        \vspace{-0.3cm}
\end{figure*}

Figure~\ref{f6} investigates the impact of the number of RIS reflecting elements on the achievable SKR. Since the relative behavior of the seven access-constrained eavesdropping scenarios has already been established in Figure~\ref{f5}, only the three fundamental single-segment access cases, namely $\mathcal{A}_E=\{\{d\}, \{t\}, \{r\}\}$ are studied here, because Figure~\ref{f5} reveals that whenever the Alice-RIS segment $\mathbf{H}_t$ is included in Eve's access set, the resulting SKR closely follows the behavior of the ${t}$ case, indicating that the Alice-RIS segment is the dominant factor governing the secrecy degradation. Conversely, the pairwise access case $\{d,r\}$ exhibits performance close to the direct-link access case ${d}$. Therefore, studying the dependence on the RIS size for the three single-segment cases adequately captures the overall behavior of all seven access scenarios while avoiding redundant numerical results. Figure~\ref{f6} presents the SKR as a function of the transmission distance for different numbers of RIS reflecting elements, $K=\{9,25,64,100,225\}$, with an $8\times8$ MIMO configuration and optimized RIS phase shifts. It is observed that the SKR decreases monotonically with increasing distance, and increasing the RIS size has little influence when Eve accesses either the direct segment $\mathbf{H}_d$ or the RIS-Bob segment $\mathbf{H}_r$, where the SKR curves for different values of $K$ almost overlap. In contrast, when Eve accesses the Alice-RIS segment $\mathbf{H}_t$, increasing the number of reflecting elements provides a significant improvement in SKR and noticeably extends the secure communication range. This behavior indicates that the passive beamforming gain introduced by a larger RIS primarily strengthens the Alice-RIS transmission stage, making the system performance considerably more sensitive to attacks on this segment than to attacks on the other propagation segments. Consequently, the influence of the RIS size on the overall secrecy performance is fundamentally governed by the characteristics of the Alice-RIS link.
%%%%%%%%%%%%%%%%%%%%%%%%%%%%%%%%%%%%%%%%%%%%%%%%%%%%%%%%
\begin{figure}[!t]
    \centering
    \includegraphics[height=6cm,width=8cm]{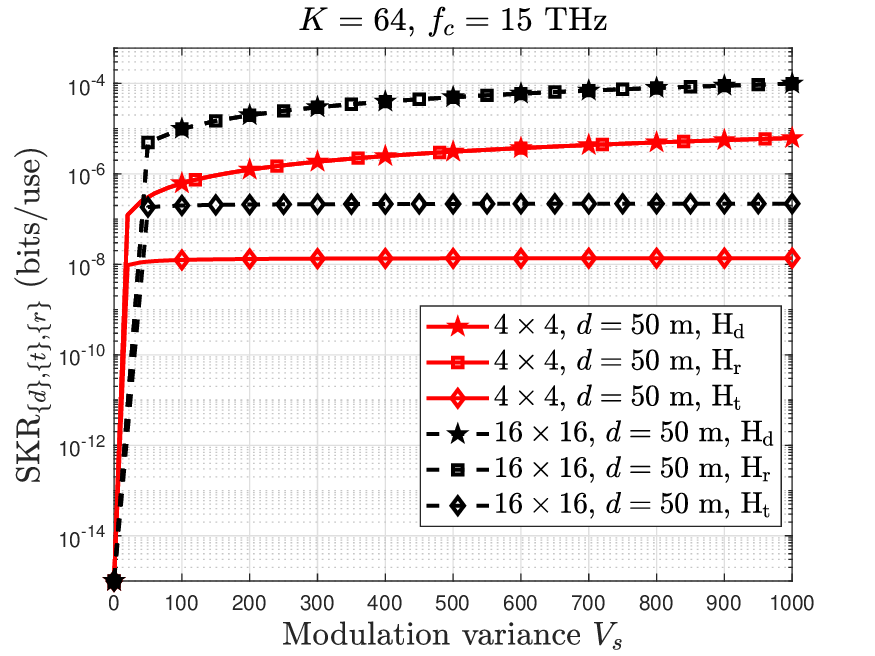}
    \caption{SKR versus modulation variance $V_s$ for $N_T=N_R=\{4, 16\}$, $f_c=15$ THz, and $d=50$ m for Eve accesses environmental modes of $\mathbf{H}_\text{d}$, $\mathbf{H}_\text{t}$, and $\mathbf{H}_\text{r}$.}
    \label{f7}
\end{figure}

Figure~\ref{f7} illustrates the SKR as a function of the modulation variance $V_s$ for $4\times4$ and $16\times16$ MIMO configurations at a transmission distance of $50\,\mathrm{m}$. As $V_s$ increases, the SKR initially improves and then gradually saturates because the additional information available to Bob is increasingly balanced by the corresponding increase in Eve's Holevo information. It is observed that among the three representative access-constrained scenarios, Eve's access to the Alice-RIS segment $\mathbf{H}_\text{t}$ yields the lowest SKR and reaches saturation much earlier than the other cases, indicating that increasing the modulation variance provides little additional secrecy once this segment is compromised. In contrast, the $\mathbf{H}_\text{d}$ and $\mathbf{H}_\text{r}$ cases continue to benefit from larger modulation variances before approaching saturation. This observation further corroborates the earlier finding that the Alice-RIS segment is the most vulnerable propagation segment from a secrecy perspective.
%%%%%%%%%%%%%%%%%%%%%%%%%%%%%%%%%%%%%%%%%%%%%%%%%%%%%%%%
\begin{figure}[!t]
    \centering
    \includegraphics[height=6cm,width=8cm]{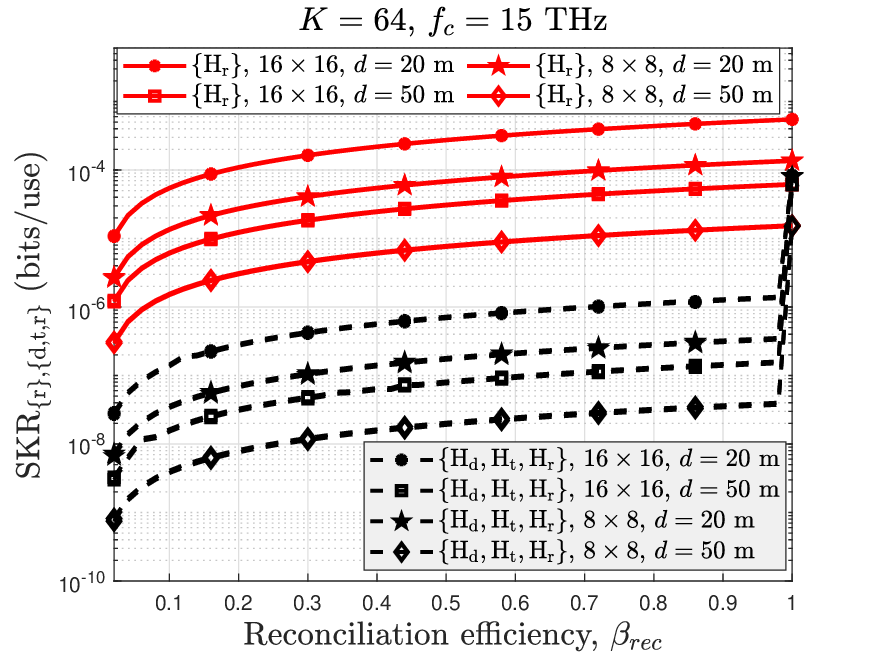}
    \caption{SKR versus reconciliation efficiency $\beta_{\mathrm{rec}}$ for $\mathcal{A}_E=\{\{r\},\{d, t, r\}\}$, considering $N_R=N_T=\{8,16\}$, $K=64$, $f_c=15$~THz, and at $d=\{20, 50\}$~m.}
    \label{f8}
\end{figure}

Figure~\ref{f8} illustrates the variation of $\text{SKR}$ with the reconciliation efficiency  $\beta_{\mathrm{rec}}$ for the best-case $\mathcal{A}_E={\{r\}}$ and the worst-case $\mathcal{A}_E={\{d, t, r\}}$ access constrained scenario for $N_t=N=\{8,16\}$ at transmission distances of $20\,\mathrm{m}$ and $50\,\mathrm{m}$. It is observed that the SKR increases monotonically with $\beta_{rec}$ for all configurations, since a higher $\beta_{rec}$ enables Bob to recover a larger fraction of the mutual information shared with Alice. A clear performance gap is observed between the two access scenarios. When Eve accesses only the RIS-Bob environmental modes $\left(\mathcal{A}_E={\{r\}}\right)$, the system achieves the highest SKR across the entire range of $\beta_{rec}$. In contrast, when Eve accesses $\mathcal{A}_E={\{d, t, r\}}$, the SKR is reduced by several orders of magnitude, reflecting the increased Holevo information available to Eve, which corroborates the observations in Figure~\ref {f5} that the secrecy performance is primarily limited by Eve's access to the Alice-RIS propagation segment, whereas access confined to the RIS-Bob segment has a comparatively smaller impact on the achievable SKR. Furthermore, larger MIMO configurations achieve higher SKRs across the entire operating range owing to their enhanced spatial degrees of freedom. Moreover, an abrupt increase in the SKR of the $\mathcal{A}_E={\{d, t, r\}}$ is observed as $\beta_{rec}$ approaches unity. This is because a highly efficient reconciliation process enables Bob to recover sufficient information to overcome Eve's advantage, leading to a rapid increase in the achievable SKR.
%%%%%%%%%%%%%%%%%%%%%%%%%%%%%%%%%%%%%%%%%%%%%%%%
\begin{figure*}
     \centering
     \begin{subfigure}[b]{0.45\textwidth}
         \centering
        \includegraphics[width=\textwidth]{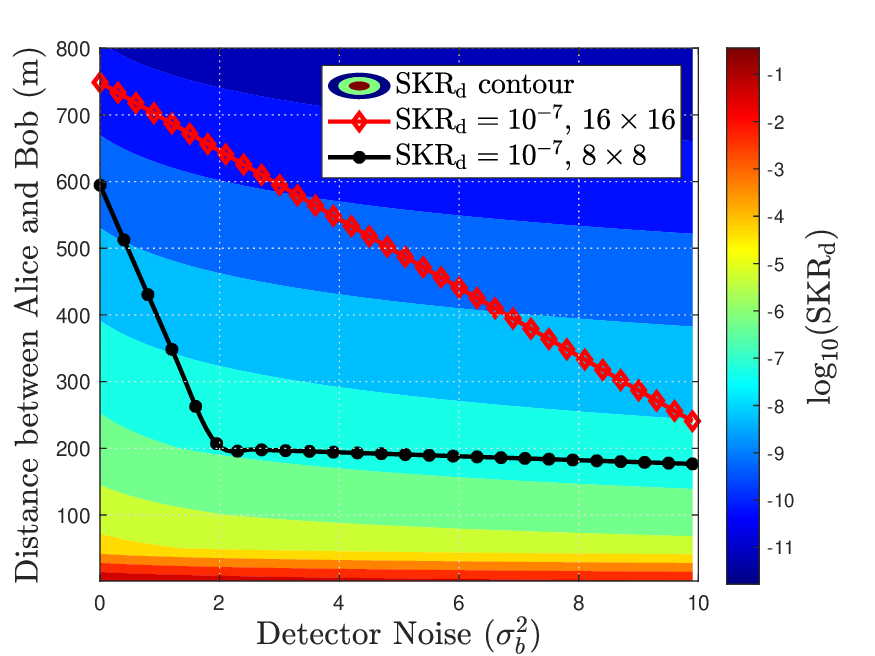}
        \caption{$\mathbf{H}_\text{d}$ segment, $\mathcal{A}_E=\{d\}$}
         \label{fig;9a}
     \end{subfigure}
     %\hspace{-0.3cm}
     \begin{subfigure}[b]{0.45\textwidth}
         \centering
        \includegraphics[width=\textwidth]{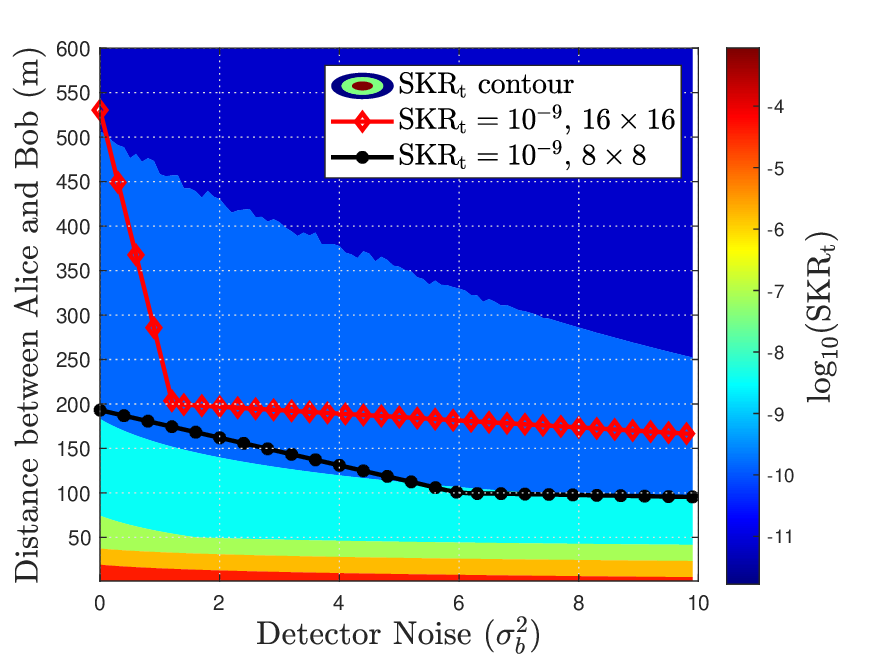}
        \caption{$\mathbf{H}_\text{t}$ segment, $\mathcal{A}_E=\{t\}$}
         \label{fig;9b}
     \end{subfigure}
     %\hspace{-0.3cm}
     \begin{subfigure}[b]{0.45\textwidth}
         \centering
        \includegraphics[width=\textwidth]{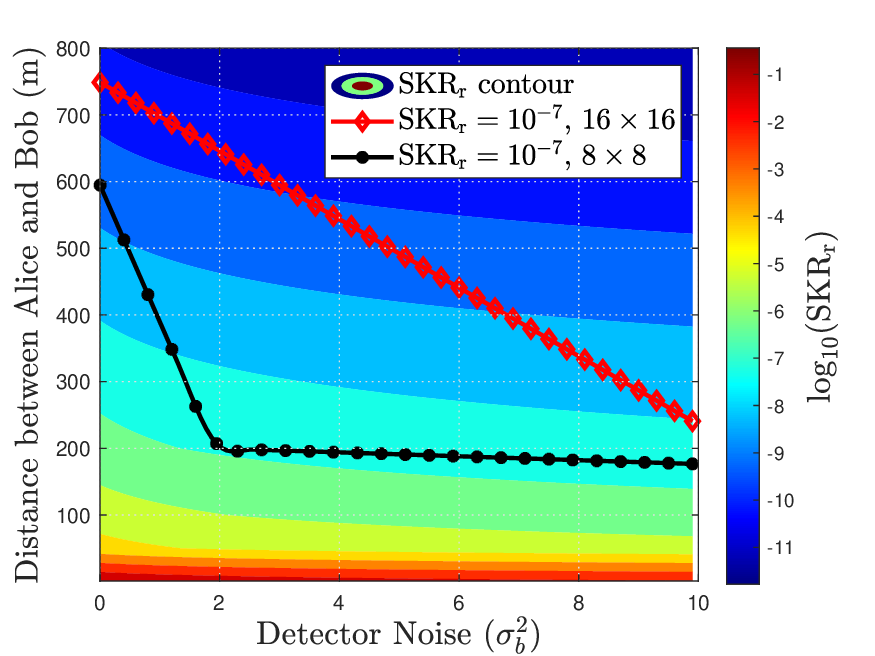}
         \caption{$\mathbf{H}_\text{r}$ segment, $\mathcal{A}_E=\{r\}$}
         \label{fig;9c}
     \end{subfigure}
     %\hspace{-0.3cm}
     \begin{subfigure}[b]{0.45\textwidth}
         \centering
        \includegraphics[width=\textwidth]{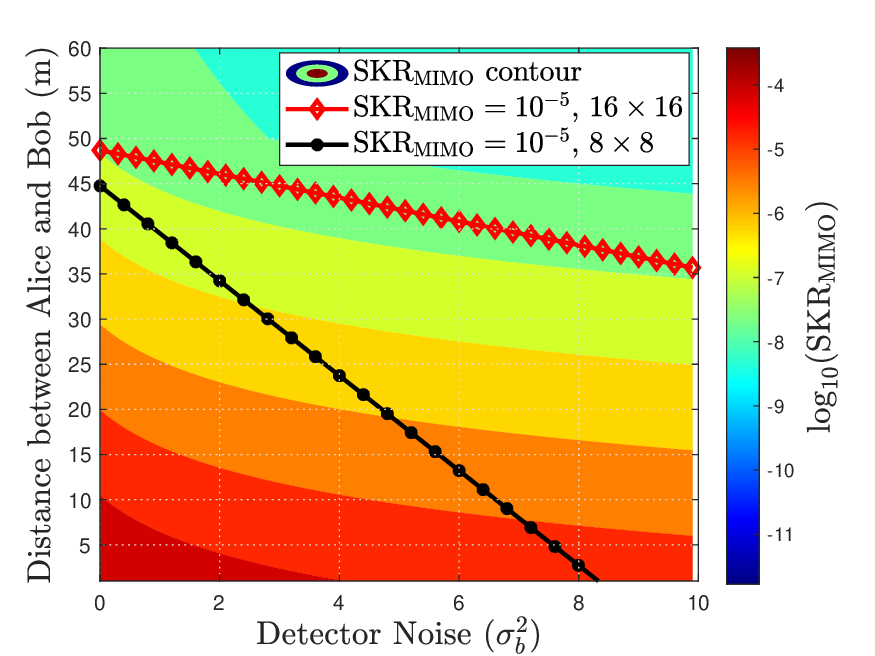}
         \caption{Eve accesses the purification of the effective channel $\mathbf{H}$}
         \label{fig;9d}
     \end{subfigure}
        \caption{Distance between Alice and Bob versus detector noise $\sigma_b^2$ for $N_{R}=N_{T}= 8,16$, $K = 64$,  $f_c$ = 15 THz, and $d_a=0.5\lambda_c$ for Eve accesses environmental modes of (a) $\mathbf{H}_\text{d}$, (b) $\mathbf{H}_\text{t}$, (c) $\mathbf{H}_\text{r}$, and (d) $\mathbf{H}$.}
        \label{f9}
        \vspace{-0.3cm}
\end{figure*}

Figure~\ref{f9} depicts the maximum secure communication distance as a function of the detector noise variance $\sigma_b^2$ for the representative access-constrained eavesdropping scenarios, namely $\mathcal{A}_E=\{\{d\}, \{t\}, \{r\}\}$, together with the benchmark global-purification attack. The results are presented for $N_R=N_T=\{8,16\}$, with $K=64$ and $f_c=15$,THz. The contour maps illustrate the variation of the achievable SKR over the detector noise variance and transmission distance, while the superimposed curves indicate the maximum communication distance corresponding to fixed SKR thresholds. It is observed that the maximum secure communication distance decreases monotonically with increasing detector noise variance for all considered eavesdropping scenarios, owing to the reduction in the mutual information between Alice and Bob. Moreover, increasing the MIMO dimension from $8\times8$ to $16\times16$ significantly enlarges the secure operating region, demonstrating the robustness offered by larger antenna arrays against receiver noise. Among the access-constrained eavesdropping scenarios shown in Figs.~\ref{fig;9a}-\ref{fig;9c}, the Alice-RIS access case, $\mathcal{A}_E=\{t\}$, yields the shortest secure communication distance, whereas the direct-link and RIS-Bob access cases, $\mathcal{A}_E=\{\{d\}, \{r\}\}$, exhibit comparatively larger secure operating regions with similar performance. Consequently, Eve's access to the Alice-RIS segment imposes the most stringent limitation on the achievable distance for secure communication. Figure~\ref{fig;9d} represents the global purification benchmark and shows a smaller secure operating region than all access-constrained eavesdropping scenarios.
%%%%%%%%%%%%%%%%%%%%%%%%%%%%%%%%%%%%%%%%%%%%%%%%%%%%%%%%%%%%%%%%%%%%%%%%%%%%%%%%%%%%%%%%%%%%%%%%%
%%%%%%%%%%%%%%%%%%%%%%%%%%%%%%%%%%%%%%%%%%%%%%%%%%%%%%%%%%%%%%%%%%%%%%%%%%%%%%%%%%%%%%%%%%%%%%%%%
%%%%%%%%%%%%%%%%%%%%%%%%%%%%%%%%%%%%%%%%%%%%%%%%%%%%%%%%%%%%%%%%%%%%%%%%%%%%%%%%%%%%%%%%%%%%%%%%%
\section{Conclusion}
In this paper, the SKR performance of an RIS-assisted THz MIMO CV-QKD system was investigated under a practical access-constrained eavesdropping model, in which Eve was assumed to access only the environmental modes associated with physically accessible propagation segments. A unified analytical framework was developed to derive closed-form SKR expressions for all single-mode, pairwise-mode, and full-segment-mode access scenarios under collective Gaussian entangling-cloner attacks with homodyne detection and reverse reconciliation. The conventional full-channel purification attack was also considered as a benchmark.
The numerical results demonstrated that the achievable SKR strongly depends on the propagation segments accessible to Eve. In particular, compromising the Alice-RIS propagation segment results in the largest degradation in secrecy performance, whereas access to the direct and RIS-Bob segments has a comparatively smaller impact. Furthermore, increasing the MIMO dimension and optimizing the RIS phase configuration significantly improve the achievable SKR and extend the secure communication range. The results also show that the proposed access-constrained model provides a less conservative and more realistic secrecy assessment than the conventional full-channel purification assumption.
Overall, the presented framework establishes RIS-assisted THz MIMO CV-QKD as a promising architecture for future secure wireless networks by explicitly accounting for the eavesdropper's physical accessibility. Future work will consider imperfect CSI, atmospheric turbulence, and practical RIS hardware impairments, including discrete phase quantization, reflection losses, and non-ideal reflecting elements.
%%%%%%%%%%%%%%%%%%%%%%%%%%%%%%%%%%%%%%%%%%%%%%%%%%%%%%%%%%%%%%%%%%%%%%%%%%%%%%%%%%%%%%%%%%%%%%%%%%%%%%%%%%%%%%%%%%%%%%%%%%%%%%%%%%%%%%%%%%%%%%%%%%%%
%%%%%%%%%%%%%%%%%%%%%%%%%%%%%%%%%%%%%%%%%%%%%%%%%%%%%%%%%%%%%%%%%%%%%%%%%%%%%%%%%%%%%%%%%%%%%%%%%%%%%%%%%%%%%%%%%%%%%%%%%%%%%%%%%%%%%%%%%%%%%%%%%%%%%%%%%%%%%%%%%%%%%%%%%%%%%%%%%%%%%%%%%%%%%%%%%%%%%%%%%%%%%%%%%%%%%%%%%%%%%%
\bibliographystyle{IEEEtran}
\bibliography{IEEEabrv,bibliography}
\end{document}